\documentclass[fleqn,usenatbib]{mnras}

\usepackage{newtxtext,newtxmath}

\usepackage[T1]{fontenc}

\DeclareRobustCommand{\VAN}[3]{#2}
\let\VANthebibliography\thebibliography
\def\thebibliography{\DeclareRobustCommand{\VAN}[3]{##3}\VANthebibliography}

\usepackage{graphicx}	
\usepackage{amsmath}	
\usepackage{pdflscape}
\usepackage{placeins}

\newcommand{\lya}{Ly$\alpha$}	
\newcommand{\kms}{~km~s$^{-1}$} 

\title[Lyman continuum escape in MUSE LAEs at $z \gtrsim 3$]{Linking UV spectral properties of MUSE \lya{} emitters at $z \gtrsim 3$ to Lyman continuum escape}

\author[Kramarenko et al.]{
I. G. Kramarenko$^{1,2}$\thanks{E-mail: im.kramarenko@gmail.com},
J. Kerutt$^{3,1}$,
A. Verhamme$^{1,4}$,
P. A. Oesch$^{1,5}$,
L. Barrufet$^{1}$,
J. Matthee$^{6,2}$,
\newauthor
H. Kusakabe$^{1}$,
I. Goovaerts$^{7}$
and T. T. Thai$^{7,8}$
\\
$^{1}$Department of Astronomy, Université de Genève, Chemin Pegasi 51, 1290 Versoix, Switzerland\\
$^{2}$Institute of Science and Technology Austria (ISTA), Am Campus 1, 3400 Klosterneuburg, Austria\\
$^{3}$Kapteyn Astronomical Institute, University of Groningen, P.O. Box 800, 9700 AV Groningen, The Netherlands\\
$^{4}$Univ. Lyon, Univ. Lyon 1, ENS de Lyon, CNRS, Centre de Recherche Astrophysique de Lyon UMR5574, 69230 Saint-Genis-Laval, France\\
$^{5}$Cosmic Dawn Center (DAWN), Niels Bohr Institute, University of Copenhagen, Jagtvej 128, København N, DK-2200, Denmark\\
$^{6}$Department of Physics, ETH Z{\"u}rich, Wolfgang-Pauli-Strasse 27, Z{\"u}rich, 8093, Switzerland\\
$^{7}$Aix Marseille Université, CNRS, CNES, LAM (Laboratoire d’Astrophysique de Marseille), UMR 7326, IPhU, 13388 Marseille, France\\
$^{8}$Department of Astrophysics, Vietnam National Space Center, Vietnam Academy of Science and Technology, 18, Hoang Quoc Viet, Nghia Do, Cau Giay,\\Ha Noi, Vietnam
}

\date{Accepted 2023 December 12. Received 2023 November 16; in original form 2023 May 12}

\pubyear{2022}

\begin{document}
\label{firstpage}
\pagerange{\pageref{firstpage}--\pageref{lastpage}}
\maketitle

\begin{abstract}
The physical conditions giving rise to high escape fractions of ionising radiation (LyC $f_{\text{esc}}$) in star-forming galaxies -- most likely protagonists of cosmic reionisation -- are not yet fully understood. Using the VLT/MUSE observations of $\sim 1400$ Lyman-$\alpha$ emitters at $2.9 < z < 6.7$, we compare stacked rest-frame UV spectra of candidates for LyC leakers and non-leakers selected based on their \lya{} profiles. We find that the stacks of potential LyC leakers, i.e. galaxies with narrow, symmetric \lya{} profiles with small peak separation, generally show (i) strong nebular \ion{O}{III]}$\lambda$1666, \ion{[Si}{III]}$\lambda$1883 and \ion{[C}{III]}$\lambda$1907$+$\ion{C}{III]}$\lambda$1909 emission, indicating a high ionisation state of the interstellar medium (ISM); (ii) high equivalent widths of \ion{He}{ii}$\lambda$1640 ($\sim1-3$~\r{A}), suggesting the presence of hard ionising radiation fields; (iii) \ion{Si}{ii}$^{*}\lambda$1533 emission, revealing substantial amounts of neutral hydrogen off the line of sight; (iv) high \ion{C}{IV}$\lambda\lambda$1548,1550 to \ion{[C}{III]}$\lambda$1907$+$\ion{C}{III]}$\lambda$1909 ratios ($\ion{C}{IV}/\ion{C}{III]}\gtrsim0.75$), signalling the presence of low-column-density channels in the ISM. In contrast, the stacks with broad, asymmetric \lya{} profiles with large peak separation show weak nebular emission lines, low \ion{He}{ii}$\lambda$1640 equivalent widths ($\lesssim 1$~\r{A}), and low $\ion{C}{IV}/\ion{C}{III]}$ ($\lesssim 0.25$), implying low ionisation states and high neutral hydrogen column densities. Our results suggest that $\ion{C}{IV}/\ion{C}{III]}$ might be sensitive to the physical conditions that govern LyC photon escape, providing a promising tool for identification of ionising sources among star-forming galaxies in the epoch of reionisation.
\end{abstract}




\section{Introduction}
\label{sec:intro}

    Ultraviolet and X-ray radiation emitted from the first stars and galaxies likely ionised the neutral hydrogen in the intergalactic medium (IGM) between $z \approx 12$ \citep[e.g.][]{Hinshaw2013,Collaboration2016} and $z \approx 6$ \citep[e.g.][]{Fan2006,McGreer2015}. This last phase transition of the Universe is known as the Epoch of Reionisation (EoR). The hydrogen reionisation could have been potentially powered by active galactic nuclei (AGNs) due to their brightness and high escape fractions of ionising radiation ($f_{\text{esc}}$). However, the AGN number density is likely to be insufficient to maintain this process at $z \gtrsim 6$ \citep[e.g][]{Parsa2018,Kulkarni2019,Shen2020}. It is currently believed that the most promising candidates for the sources of the ionising photons are young massive stars in star-forming galaxies (SFGs), although their contribution is still unclear, primarily due to the uncertainties associated with $f_{\text{esc}}$.
    
    Recent studies have shown that $f_{\text{esc}}$ in SFGs at $z \gtrsim 6$ has to be on average at least 10--20~per~cent to successfully reproduce the observed constraints on the reionisation history \citep[e.g.][]{Ouchi2009,Robertson2015,Bouwens2015,Khaire2016,Naidu2020}.
    Direct measurements of $f_{\text{esc}}$, however, are severely hampered at such redshifts because ionising radiation, or Lyman continuum (LyC, $\lambda < 912$~\AA), is absorbed by the neutral hydrogen in the IGM \citep[e.g.][]{Madau1995,Inoue2014}. The current solution to this problem lies in the studies of LyC-leaking SFGs in the local Universe \citep[e.g.][]{Izotov2016,Izotov2016_1,Izotov2018_1,Izotov2018,Schaerer2018,Wang2019,Flury2022,Xu2022} and at intermediate redshifts \citep[$1 \lesssim z \lesssim 4$; e.g.][]{Vanzella2015,Shapley2016,Vanzella2018,Rivera-Thorsen2017,Marques-Chaves2021,Saxena2022a}, often accompanied by spectral stacking with the aim of obtaining higher signal-to-noise (S/N) data and averaging out the spatial variations of the IGM transmission \citep[e.g.][]{Marchi2017,Steidel2018,Mestric2021}. The low-$z$ observations help to develop indirect tracers of $f_{\text{esc}}$ which can then be extrapolated to the galaxies in the EoR.
    
    Historically, one of the first proposed indicators of LyC leakage was the \ion{[O}{iii]}$\lambda\lambda$4959,5007 to \ion{[O}{ii]}$\lambda\lambda$3726,3729 ratio \citetext{O$_{32}$; \citealp{Jaskot2013,Nakajima2014}}. Recent observations suggest, however, that the correlation between O$_{32}$ and $f_{\text{esc}}$ is weak \citep[e.g.][]{Izotov2018,Naidu2018}, possibly due to the complex geometry and the kinematics of the interstellar medium (ISM) modulating LyC escape \citep{Bassett2019,McKinney2019,Nakajima2020}. A better proxy for $f_{\text{esc}}$ might be an ensemble of UV low-ionisation state (LIS) absorption lines (e.g. \ion{Si}{ii}$\lambda$1260, \ion{O}{i}$\lambda$1302 and \ion{C}{ii}$\lambda$1334). Observationally, the neutral gas covering fraction ($f_{\text{cov}}$) inferred from the LIS lines correlates with the fraction of LyC photons escaping through low-column-density channels in the ISM \citep[e.g.][]{Reddy2016,Gazagnes2018,Chisholm2018,Saldana-Lopez2022}, although simulations demonstrate a substantial scatter in this relation \citep{Mauerhofer2021}. Alternatively, the ISM conditions driving LyC escape could be probed with resonant nebular emission lines sensitive to the radiation transfer effects. These include the \ion{Mg}{ii}$\lambda\lambda$2796,2803 doublet \citep{Feltre2018,Henry2018,Chisholm2020}, the \ion{C}{IV}$\lambda\lambda$1548,1550 doublet \citep{Schaerer2022,Saxena2022} and the Lyman-alpha (\lya{}) line.
    
    The \lya{}-LyC relationship has been extensively explored in the literature. \citet{Verhamme2015} studied \lya{} and LyC escape in the two following configurations of the ISM: (i) density-bound nebulae, and (ii) ionisation-bounded nebulae with holes, also referred to as the ``picket-fence'' model. They found that in the first case the \lya{} line has a narrow profile and a small velocity offset with respect to systemic redshift ($\upsilon^{\text{red}}_{\text{peak}} \lesssim 150$\kms{}), whereas in the second case the \lya{} line is at systemic redshift. They proposed to use small \lya{} peak separation ($\upsilon_{\text{sep}} \lesssim 300 $\kms{}) for double-peaked \lya{} profiles as an indicator of low neutral hydrogen column density and thus potentially high $f_{\text{esc}}$. \citet{Dijkstra2016} carried out radiative transfer simulations of the clumpy ISM and found that the LyC-emitting galaxies have narrower, more symmetric \lya{} line profiles. \citet{Kimm2019} and \citet{Kakiichi2021} obtained similar results from radiation-hydrodynamic simulations of turbulent molecular clouds. Finally, recent observational studies have shown that local SFGs with high $f_{\text{esc}}$ typically have a narrow \lya{} line with small peak separation, confirming the theoretical expectations \citep{Izotov2016,Verhamme2017,Izotov2018,Izotov2021,Izotov2022,Flury2022}.
    
    These results demonstrate that the \lya{} line profile could provide robust predictions for $f_{\text{esc}}$, potentially even at high redshift. \lya{} profiles have been already used to select candidates for LyC leakers and non-leakers from a representative sample of Lyman-alpha emitters (LAEs) at $z \approx 2$ \citep{Naidu2022}. In this paper, we apply the same technique to examine the possible relationship between LyC escape and rest-frame UV spectral properties of $\sim$1400~LAEs observed with the Multi Unit Spectroscopic Explorer \citetext{MUSE; \citealp{Bacon2010}} at $2.9 < z < 6.7$. We combine the 3D~spectroscopic data from the MUSE-Wide \citep{Urrutia2019} and the MUSE \textit{Hubble} Ultra Deep Field \citetext{MUSE HUDF; \citealp{Bacon2017}} surveys and select a sample of LAEs with reliable detections of the \lya{} line (\autoref{sec:data}). In \autoref{sec:subsamples}, we use the observed properties of the \lya{} line profile (e.g. peak separation) to select groups of LAEs with potentially different $f_{\text{esc}}$. In \autoref{sec:stacks}, we stack the individual spectra of our galaxies to obtain high S/N detections of the rest-frame UV lines, namely, high-ionisation nebular emission lines (e.g. \ion{C}{IV}$\lambda\lambda$1548,1550, \ion{He}{ii}$\lambda$1640 and \ion{[C}{III]}$\lambda$1907$+$\ion{C}{III]}$\lambda$1909), ISM absorption lines (e.g. \ion{O}{i}$\lambda$1302$+$\ion{Si}{II}$\lambda$1304 and \ion{C}{ii}$\lambda$1334) and fine-structure emission lines (e.g. \ion{Si}{II}$^{*}\lambda$1533). In \autoref{sec:results}, we compare the stacked spectra of LyC-leaker and non-leaker candidates to examine the ionising properties of galaxies and the physical conditions in their ISM indicative of high $f_{\text{esc}}$. In \autoref{sec:summary}, we present the summary of our main findings and discuss the implications of our work for reionisation studies.

\section{Data and sample}
\label{sec:data}

    \subsection{Spectroscopic data from MUSE}
    For the purpose of this study, we use the spectroscopic data from the MUSE-Wide \citep{Urrutia2019} and MUSE~HUDF \citetext{\citealp{Bacon2017}; Data Release I} surveys taken during the guaranteed time observations (GTOs) of the MUSE consortium. The MUSE integral-field spectrograph is a powerful tool for spectroscopic diagnostics of emission-line galaxies thanks to a combination of a large field of view ($1\times 1$~arcmin$^2$), high resolving power (ranging from $R \approx 1800$ in the blue to $\approx 4000$ in the red) and a large simultaneous spectral range (4750--9350~\r{A}; \citealp{Bacon2015}). The GTO programs take full advantage of these unique spectroscopic capabilities of the MUSE instrument, enabling studies of large, un-targeted samples of faint ($\text{M}_{\text{UV}}$ down to $\approx -16$) LAEs at $z > 3$.
    
    Being different in depths and sizes, MUSE-Wide and MUSE HUDF effectively complement each other, probing the \lya{}-bright ($\mathrm{L}_{\mathrm{Ly}\alpha} \gtrsim \mathrm{L}^*$) and faint ($\mathrm{L}_{\mathrm{Ly}\alpha} \lesssim 0.1\,\mathrm{L}^*$) populations of LAEs, respectively. MUSE-Wide, a relatively wide and shallow component of the GTO ``wedding cake'', covers 100 $1\times1$~arcmin$^2$ fields at one hour observation time, among which 60 are in the CANDELS \citep{Koekemoer2011, Grogin2011}/GOODS-South \citetext{GOODS-S; \citealp{Giavalisco2004}} field, and 23 are in the CANDELS/COSMOS \citep{Scoville2007} field. The other eight pointings of MUSE-Wide are in the HUDF parallel fields. MUSE HUDF, the pencil-beam component of the GTO ``wedding cake'', targets a much smaller area but for longer exposure times. This survey consists of nine fields of ten hours depth in HUDF, completing the 100~fields of MUSE-Wide, plus a single UDF-10 field with 31 hours exposure time.
    
    In this paper, we use the 1D~spectra extracted from the reduced MUSE-Wide and MUSE HUDF data cubes by \citet{Schmidt2021}. The untargeted search for emission line sources in the MUSE data cubes was performed using the \textsc{LSDCat} tool \citep{Herenz2017}. Each of the detected sources was then assigned a subjective confidence C between zero (lowest confidence) and three (highest confidence) depending on the uncertainty of the line classification. The latter was carried out using the custom graphical user interface \textsc{QtClassify} \citep{Kerutt2017}. Finally, the optimally extracted spectra were obtained with \textsc{TDOSE} \citep{Schmidt2019} using morphological \textit{Hubble} Space Telescope (HST) models as templates. To preserve the self-consistency of this procedure, we decided against incorporating the data from the MUSE eXtremely Deep Field (MXDF) survey added to the second data release of MUSE HUDF \citep{Bacon2023}.
    
    \subsection{Sample selection}
    We select MUSE LAEs for our analysis as follows. First, we only consider LAEs with a confidence $\text{C}\geq2$, i.e.~sources in which \lya{} is detected at $\text{S/N} > 5$ and the line profile is compatible with typical \lya{} shapes \citetext{see \citealp{Bacon2023} for a detailed description of MUSE confidence levels}. Next, we note that some spectra might represent the same galaxy in the total LAE sample due to a partial overlap between the MUSE-Wide fields, the UDF mosaic and the UDF-10 field. Among these duplicates, we consider spectra with the highest observation time only. Finally, we exclude 19 superpositions (spatially overlapping sources at different redshifts) and three galaxies classified as AGN in the 7 Ms \textit{Chandra} Deep Field-South Survey catalogs \citep{Luo2017}.
    
    Our final sample amounts to 1422~LAEs in the redshift range $2.92 < z < 6.64$, with 697 objects from the MUSE-Wide survey and 725 objects from the MUSE HUDF survey. Unlike \citet{Feltre2020}, who carried out a similar spectroscopic analysis of MUSE LAEs at $z > 3$, we include in our sample objects at redshifts where the MUSE spectral range does not cover several important rest-frame UV lines (e.g. the \ion{[C}{III]}$\lambda$1907$+$\ion{C}{III]}$\lambda$1909 doublet at $z > 3.9$). We find that the evolution of redshift coverage with wavelength does not affect our conclusions (see \autoref{sec:stacks}).
    
    \subsection{Sample properties}
\begin{figure}
        \includegraphics[width=\linewidth]{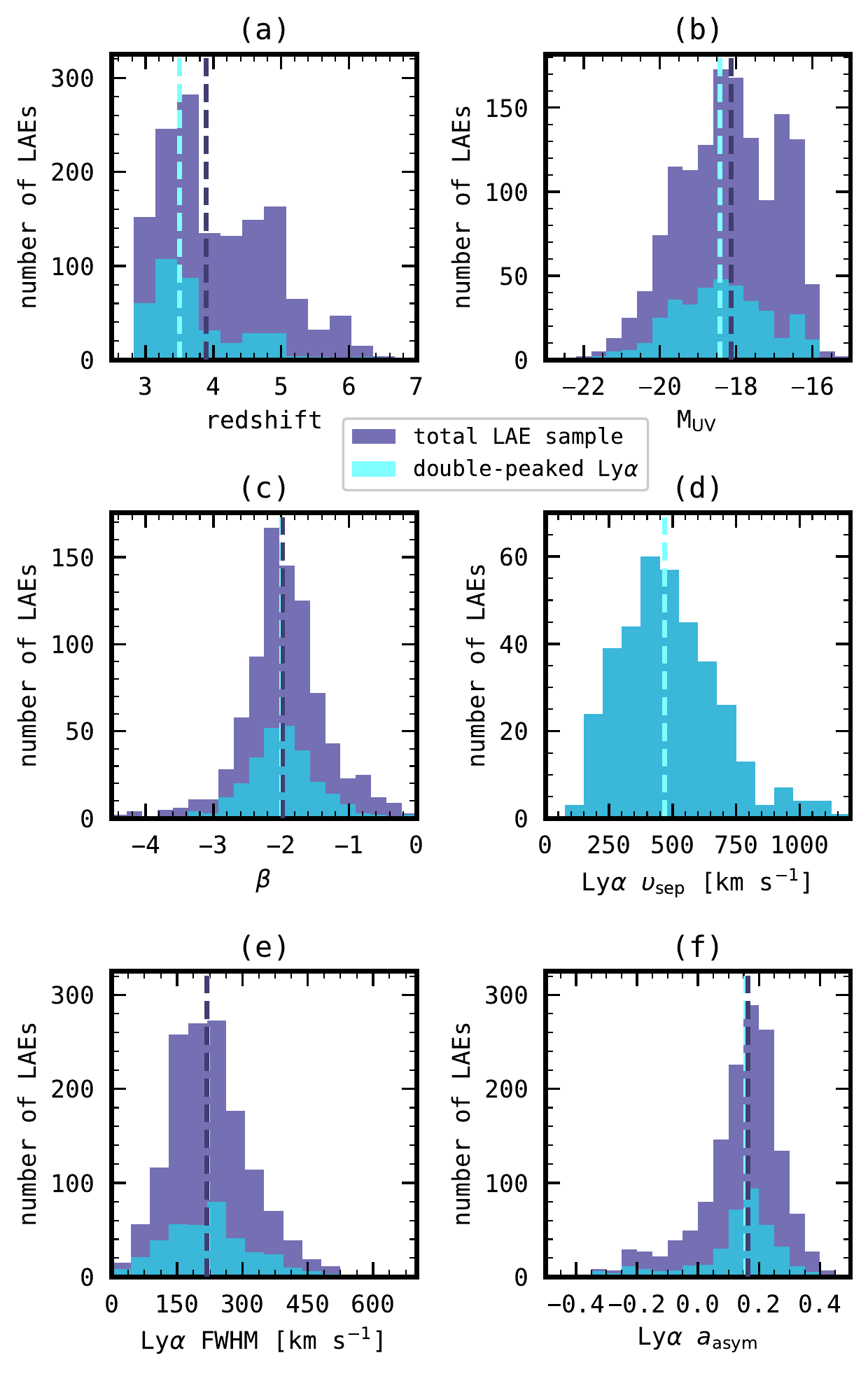}
        \caption{Observed properties of LAEs studied in this work \citetext{data from \citealp{Kerutt2022}}. The distributions are shown for the total sample (1422 LAEs; dark blue) and the sample of galaxies with double-peaked \lya{} profiles (369 LAEs; cyan). \textbf{From panel a–f:} redshift, absolute UV magnitude, UV continuum slope, \lya{}~peak separation, \lya{}~FWHM, \lya{}~asymmetry parameter. Dashed vertical lines indicate the median values of the distributions.}
        \label{fig:sample}
    \end{figure}

    The distributions of the observed properties of our LAEs are shown in \autoref{fig:sample}, with data taken from \citet{Kerutt2022}. The redshifts (\autoref{fig:sample}, panel a) are measured using the \lya{} line and have a median value of 3.89. We caution that redshift estimates based on \lya{} emission have systematic uncertainties due to resonant scattering of \lya{} photons by neutral gas in the ISM. High neutral hydrogen column density leads to a significant shift of the peak of the \lya{} line ($\upsilon^{\text{red}}_{\text{peak}}$), typically of the order of a few hundred \kms{} \citep[e.g.][]{Shapley2003,Steidel2010,Song2014,Hashimoto2015,Muzahid2020,Matthee2021}. However, having accurate redshift measurements is crucial for our study to correctly perform spectral stacking. We describe the approach that we used to recover systemic redshifts in \autoref{sec:stacks}.

    As shown in \autoref{fig:sample}, panel b, the absolute UV magnitudes of our LAEs span the range of $-21.9 < \text{M}_{\text{UV}} < -15.6$, with a median value of $-18.1$. Thanks to the inclusion of the MUSE HUDF data, our sample probes $\sim 1-2$~dex fainter magnitudes compared to the samples of LAEs selected using the narrow-band technique \citep[e.g.][]{Ouchi2008,Kashikawa2011,Zheng2014,Nakajima2018a,Matthee2021}, or samples of Lyman-break galaxies \citep[e.g.][]{Shapley2003,Stark2010}. Magnitudes as faint as $\text{M}_{\text{UV}} \sim -16$ or even fainter, down to $\text{M}_{\text{UV}} \sim -14$, have been previously achieved mostly in the studies of gravitationally lensed LAEs \citep[e.g.][]{Stark2014,Vieuville2020,Bouwens2022}.

    The distribution of the UV continuum slopes ($\beta$) is shown in \autoref{fig:sample}, panel c. These estimates are based on the HST observations of LAEs with at least two detections in the HST filter bands (856 objects, or 60~per~cent of the total sample). The fact that most of our LAEs have blue UV slopes with a median $\beta = -2.1$ suggests low dust content, as has been previously reported in other LAE studies \citep[e.g.][]{Matthee2021}. Therefore, we can safely neglect the effect of dust attenuation when analysing the spectra of our LAEs (\autoref{sec:results}).

    \subsection[Lyman-alpha line profile statistics]{\lya{} line profile statistics}
    \label{sec:lya_statistics}

    In this study, we select candidates for LyC-leakers and non-leakers using the properties of the \lya{} line profile such as the peak separation, full width at half maximum (FWHM) and asymmetry parameter. We show the distributions of the \lya{} profile measurements in \autoref{fig:sample}, panels d-f. We find that our galaxies exhibit a large variety of \lya{} profiles in terms of the number of peaks, width, and skewness \citetext{see also Figure~5 from \citealp{Kerutt2022}}. This implies a wide range of the physical conditions in the ISM that shape radiative transfer processes.

    The peak separations -- distances between the red and the blue peaks of the double-peaked \lya{} profiles -- range from $\approx100$~\kms{} to $\approx1200$~\kms{}, with a median value of $\approx470$~\kms{} (\autoref{fig:sample}, panel d). We note that only 369 LAEs, or 26~per~cent of the total sample, have resolved double-peaked \lya{} profiles. By comparing the UV properties of these galaxies with that of the total sample, we find that the double-peaked \lya{} sample is representative of the parent population (dark blue and cyan histograms in \autoref{fig:sample}).
    
    There might be several explanations for the lack of the double-peaked structure in most of the observed \lya{} profiles. In some cases, low-S/N data and/or the presence of the IGM absorption hinder the detection of the blue peak (hereafter the blue bump), which is typically several times weaker than its red counterpart. Alternatively, the blue bump might be blended with the red peak due to the limited spectral resolution of the MUSE instrument, effectively resulting in a single-peaked \lya{} profile. We discuss the limitations of using the double-peaked \lya{} sample for the analysis of LyC escape in \autoref{sec:subsamples}.
    
    In addition to the \lya{} peak separation, we use the FWHM and asymmetry parameter of the red peak of the \lya{} line in our selection of LyC-leaker candidates. These properties are measured in \citet{Kerutt2022} by fitting the \lya{} profile with an asymmetric Gaussian function:
    \begin{equation}
    f(\lambda) = A \exp{\left( - \frac{(\lambda-\lambda_0)^2}{2 \sigma^2_{\text{asym}}} \right)} + f_0,
    \end{equation}
    where $A$ is the amplitude, $\lambda_0$ is the wavelength of the red peak, $\sigma_{\text{asym}}$ is the asymmetric dispersion and $f_0$ is the continuum level. The asymmetric dispersion is described by $\sigma_{\text{asym}} = a_{\text{asym}} (\lambda - \lambda_0) + d$, where $a_{\text{asym}}$ is the asymmetry parameter and $d$ is the typical width of the line. The asymmetry parameter is positive for the vast majority of the sample (87~per~cent) and has a median value of $+0.16$, indicating that the \lya{} line typically has a red wing (\autoref{fig:sample}, panel f). Negative values of $a_{\text{asym}}$ suggest the presence of a blue wing which might be the case for LAEs with an unresolved blue bump blended with the red peak. The FWHMs range from $\approx60$~\kms{} to $\approx440$~\kms{}, with a median value of $\approx220$~\kms{} (\autoref{fig:sample}, panel e). The FWHMs are corrected for the wavelength-dependent line spread function (LSF) of MUSE \citep{Bacon2017} under the assumption that the \lya{} line profile can be approximated by a Gaussian \citetext{see \citealp{Kerutt2022} for the details}.

\section{Selecting LyC-leaker candidates}
\label{sec:subsamples}

    In this section, we classify our LAEs as potential LyC leakers and non-leakers using the properties of the \lya{} line profile described in \autoref{sec:data} (peak separation, FWHM and asymmetry parameter). We briefly discuss the physics behind the relationship between the \lya{} line profile and LyC escape to motivate our selection criteria. In addition, we estimate the median $f_\text{esc}$ for each resulting group of LAEs wherever possible. At the end of this section, we examine the global UV properties ($\text{M}_{\text{UV}}$, $\beta$) of LyC-leaker and non-leaker candidates, and make a comparison with the literature to place the results of our classification in a broader context.
    
    The accuracy of our classification procedure depends on uncertainties of the observed \lya{} properties, i.e. a LAE might be attributed to a wrong group if its \lya{} peak separation, FWHM and/or asymmetry parameter have high statistical uncertainties. Thus, on the one hand, we follow a strategy of excluding LAEs whose observed \lya{} properties are highly uncertain. On the other hand, we keep the subsamples large enough to achieve high S/N values of the stacked spectra to be able to detect weaker rest-frame UV lines compared to \lya{} (\autoref{sec:stacks} and \autoref{sec:results}).
    
    \begin{figure}
        \includegraphics[width=\linewidth]{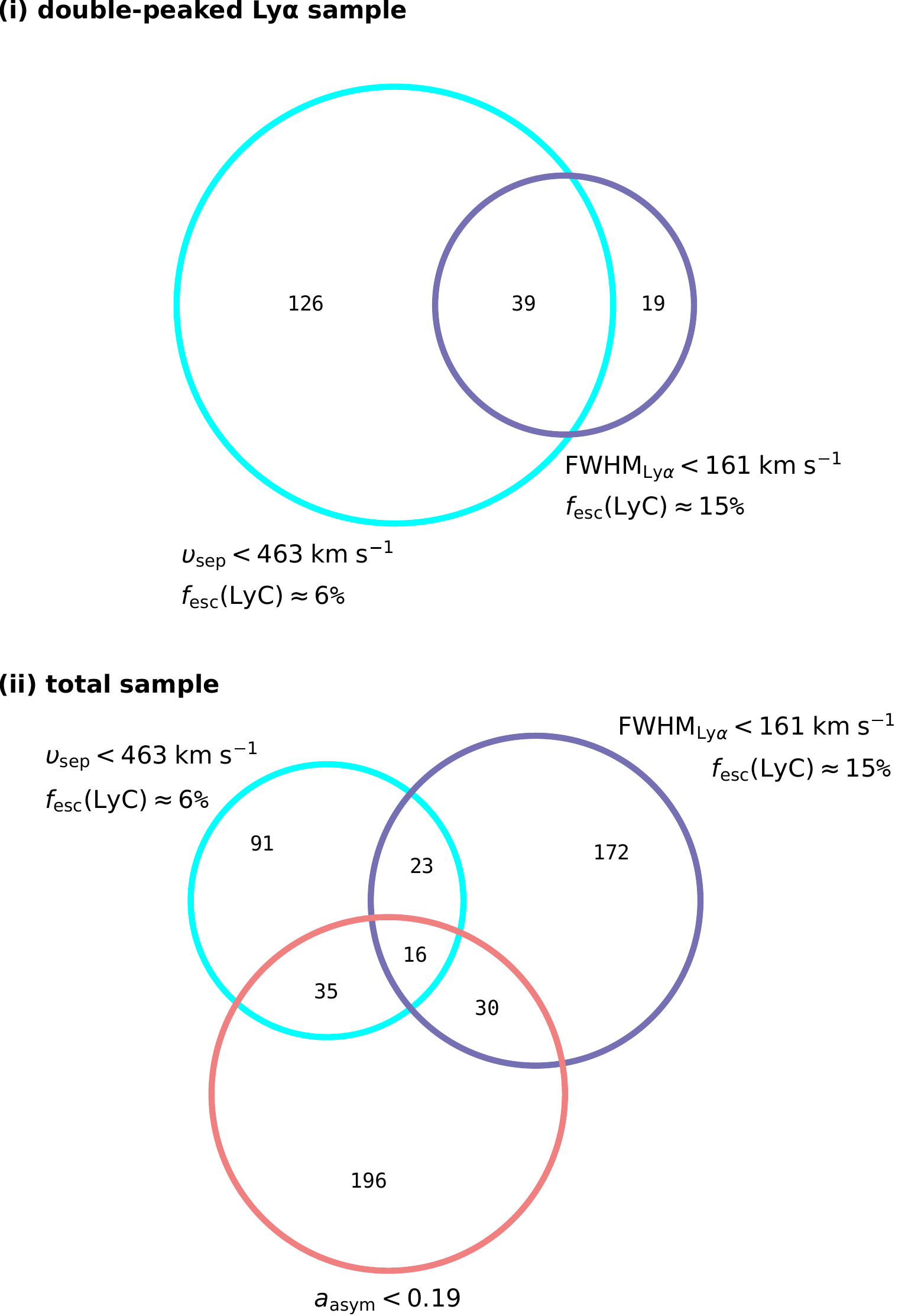}
        \caption{Venn diagram showing the samples of LyC-leaker candidates selected as having low \lya{} peak separation ($<463$~\kms{}; cyan), FWHM ($<161$~\kms{}; dark blue) and asymmetry parameter ($< 0.19$; red). \textbf{Top:} LAEs with double-peaked \lya{} profiles. \textbf{Bottom:} Total sample. The estimates of the median $f_{\text{esc}}$ are based on the fit from \citet{Izotov2018}. All three properties of the \lya{} line act as tracers of $f_{\text{esc}}$ due to their high sensitivity to the neutral hydrogen column density -- a key parameter of the ISM controlling escape of ionising photons.}
        \label{fig:leakers_venn}
    \end{figure}
    
    \subsection[Lyman-alpha peak separation]{\lya{} peak separation}
    The \lya{} peak separation is considered as one of the most reliable tracers of $f_{\text{esc}}$ in low-redshift galaxies \citep[e.g.][]{Izotov2021}. In the density-bounded scenario, small peak separation indicates that \lya{} photons escape from an almost fully ionized ISM with little scattering thanks to low column densities of neutral hydrogen \citep[e.g.][]{Verhamme2015}. The same physical conditions allow LyC photons escape from the ISM without being absorbed by neutral gas.
    
    The limited number of LAEs with resolved blue bumps (see \autoref{sec:lya_statistics}) prevents us from having more than two groups selected based on the peak separation. After excluding 40~objects with $\sigma_{\upsilon_{\text{sep}}}> 200$~\kms{}, we obtain a sample of 329 LAEs with resolved double-peaked \lya{} line profiles (23~per~cent of the total sample). From this sample, we select two groups of roughly the same size with LAEs having $\upsilon_{\text{sep}} < 463$~\kms{} (165 objects; potential leakers) and $\upsilon_{\text{sep}} > 463$~\kms{} (164 objects; potential non-leakers), respectively.
    
    \citet{Izotov2018} collected $\upsilon_\text{sep}$ measurements for a sample of low-redshift LyC-leaking galaxies and found the following empirical relationship between $f_\text{esc}$ and $\upsilon_\text{sep}$:
    \begin{equation}
    \label{eq:izotov_2}
        f_\text{esc} (\upsilon_\text{sep}) = \frac{3.23 \cdot 10^4}{\upsilon_\text{sep}^2} - \frac{1.05 \cdot 10^2}{\upsilon_\text{sep}} + 0.095,
    \end{equation}
    where $\upsilon_\text{sep}$ is in \kms{}. We use this equation to estimate the expected median LyC~$f_\text{esc}$ and obtain values of 6.0~per~cent and 1.1~per~cent for the low- and high-$\upsilon_\text{sep}$ subsamples, respectively. Therefore, our classification of a galaxy as a LyC leaker based on the \lya{} peak separation allows lower values of $f_\text{esc}$ compared to the thresholds usually adopted in other studies ($f_\text{esc} > 20$~per~cent, \citealp{Naidu2022}; or $f_\text{esc} > 10$~per~cent, \citealp{Schaerer2022}).

    We note that the \lya{} line profile can be affected by the IGM (and the CGM) absorption, especially at high redshift where the universe becomes more neutral. Taking into account the stochastic nature of this effect, we caution that the resulting $\upsilon_\text{sep}$ subsamples might be ``contaminated'' by LAEs observed along the sightlines with suppressed \lya{} transmission. Importantly, both theoretical and observational works show that the neutral hydrogen in the IGM predominantly attenuates the blue part of the \lya{} line \citep[e.g.][]{Laursen2011,Hayes2021}. This highlights the importance of other properties of the \lya{} profile (FWHM, asymmetry parameter) which are measured from the red peak of the \lya{} line and therefore less sensitive to the IGM absorption.
    
    \subsection[Lyman-alpha FWHM]{\lya{} FWHM}
    Similarly to the \lya{} peak separation, the \lya{} FWHM traces the neutral hydrogen column density in the ISM. For the density-bounded geometry, a narrower \lya{} line profile suggests less scattering of \lya{} photons, higher fractions of ionized gas in the ISM and higher $f_{\text{esc}}$. We select 964~LAEs with $\sigma_{\text{FWHM}_{\text{Ly}\alpha}} < 100$~\kms{} and split them into four FWHM$_{\text{Ly}\alpha}$ quartiles (Q1-Q4) of 241 LAEs each as follows: (Q1) $\text{FWHM}_{\text{Ly}\alpha}<161$~\kms{}, (Q2) $\text{FWHM}_{\text{Ly}\alpha} \in \left[161, 208\right] $~\kms{}, (Q3) $\text{FWHM}_{\text{Ly}\alpha} \in \left[208, 260\right] $~\kms{}, and (Q4) $\text{FWHM}_{\text{Ly}\alpha} > 260$~\kms{}. We find that 58 objects (24~per~cent) in the group of LyC-leaker candidates (Q1, the lowest FWHMs) have \lya{} line profiles with two resolved peaks. Most of these galaxies (39 LAEs, or 67~per~cent) fall into the low-$\upsilon_\text{sep}$ subsample (\autoref{fig:leakers_venn}, top), reflecting the fact that FWHM$_{\text{Ly}\alpha}$ correlates with $\upsilon_\text{sep}$ since both depend on the neutral hydrogen column density as discussed above.
    
    Expanding shell models predict that FWHM$_{\text{Ly}\alpha}$ could be approximated by half of the peak separation \citep{Verhamme2018}. This allows us to use \autoref{eq:izotov_2} once again, this time to estimate $f_\text{esc}$ for the FWHM$_{\text{Ly}\alpha}$ subsamples. We caution that the relation between FWHM$_{\text{Ly}\alpha}$ and $\upsilon_\text{sep}$ holds as long as the medium enables at least some degree of \lya{} photon scattering, otherwise the \lya{} profile would have a single peak at systemic velocity. However, this should be the case for most star-forming galaxies since the minimal column densities required for a substantial number of scattering events are extremely low (for instance, the \lya{} forest is observed at column densities as low as $\mathrm{n}_{\mathrm{HI}}\sim 10^{13}$~cm$^{-2}$).
    
    We find that the expected median $f_\text{esc}$ ranges from 1.1~per~cent in the highest FWHM$_{\text{Ly}\alpha}$ quartile (Q4) to 15~per~cent in the lowest FWHM$_{\text{Ly}\alpha}$ quartile (Q1). Therefore, the strongest leakers selected by the \lya{} FWHM are likely to have higher escape fractions than the LAEs with the smallest peak separations ($f_{\text{esc}} \approx 6$~per~cent). Nonetheless, even the lowest FWHM$_{\text{Ly}\alpha}$ quartile (Q1) has a median $f_\text{esc}$ smaller than the ones observed in the individual LyC leakers at $2 \lesssim z \lesssim 4$ \citetext{$\gtrsim 20$~per~cent; see \citealp{Izotov2021} for a review}. Most likely, we are probing lower escape fractions because our sample is free of a selection bias typical for the individual detections which favour galaxies with the highest LyC fluxes \citetext{see \citealp{Steidel2018} for a discussion}.
    
    \subsection[Lyman-alpha asymmetry parameter]{\lya{} asymmetry parameter}
    The asymmetry of the \lya{} line profile provides an alternative way of estimating the optical depth of neutral gas in the ISM \citep[e.g.][]{Erb2014}. More symmetric profiles might suggest the presence of an unresolved blue bump blended with the red peak of \lya{} (see \autoref{sec:lya_statistics}), or \lya{} emission at systemic velocity in cases where \lya{} (and LyC) photons escape through a porous ISM with little scattering \citep[e.g.][]{Verhamme2015}. However, quantifying the asymmetry of \lya{} profiles is challenging at relatively low spectral resolutions ($R \lesssim 3000$) and high skewness \citep{Childs2018}. After applying a moderate cutoff on the $a_{\text{asym}}$ uncertainty ($\sigma_{a_{\text{asym}}} < 0.07$), we are left with 554~LAEs, or 39~per~cent of the total sample. From these LAEs, we select two groups with $a_{\text{asym}} < 0.19$ (potential leakers) and $a_{\text{asym}} > 0.19$ (potential non-leakers), both comprising of 277 objects.
    
    The quantitative comparison between $a_{\text{asym}}$ and $f_{\text{esc}}$ is missing from the literature. If we use FWHM$_{\text{Ly}\alpha}$ to estimate the median $f_{\text{esc}}$, we obtain values of 2.0~per~cent (low-$a_{\text{asym}}$ subsample) and 2.8~per~cent (high-$a_{\text{asym}}$ subsample) that do not comply with our classification -- LAEs with more symmetric \lya{} profiles are expected to have higher $f_{\text{esc}}$. Furthermore, we find that only 81 objects (29~per~cent) from the low-$a_{\text{asym}}$ subsample are part of the low-$\upsilon_\text{sep}$ and -FWHM$_{\text{Ly}\alpha}$ subsamples (\autoref{fig:leakers_venn}, bottom). This could indicate that $a_{\text{asym}}$, on the one hand, and $\upsilon_\text{sep}$ and FWHM$_{\text{Ly}\alpha}$, on the other hand, are sensitive to different geometries of the ISM (namely, the ``picket-fence'' and the density-bounded geometries) which could both drive LyC~escape but not necessarily under the same physical conditions. We discuss different ISM geometries further in \autoref{sec:results} when examining the stacked spectra of our LAEs.

    \subsection{Connection to the global UV properties}

    Finally, we compare the global UV properties (M$_{\text{UV}}$, $\beta$) of potential LyC leakers and non-leakers. We find that M$_{\text{UV}}$ becomes fainter with increasing $f_{\text{esc}}$ inferred from \autoref{eq:izotov_2}, although the correlation is generally weak. The FWHM$_{\text{Ly}\alpha}$ subsamples show the largest variations in M$_{\text{UV}}$, with magnitudes ranging from M$_{\text{UV}}\approx-17.7$ for leakers (FWHM$_{\text{Ly}\alpha}$ Q1) to M$_{\text{UV}}\approx-19.0$ for non-leakers (FWHM$_{\text{Ly}\alpha}$ Q4). This trend is consistent with the cosmological radiation-hydrodynamical simulations from \citet{Rosdahl2022}, who demonstrated that $f_{\text{esc}}$ peaks at M$_{\text{UV}}\approx-17$ (see Figure~9 in their paper). Having deeper data e.g. from lensing field observations would be important to test their predictions for a declining $f_{\text{esc}}$ at magnitudes below M$_{\text{UV}}\approx-17$.
    
    We also find that on average, the LyC-leaker candidates are bluer. For instance, the low- and high-$\upsilon_\text{sep}$ subsamples have median $\beta=-2.16$ and $\beta=-1.98$, respectively. This result is in agreement with \citet{Chisholm2022} who reported a 6$\sigma$ inverse correlation between the $\beta$ slope and $f_{\text{esc}}$ based on the Low-redshift Lyman Continuum Survey (LzLCS) observations of 89~SFGs at $z \approx 0.3$. We caution, however, that the dynamic range of $\beta$ slopes across our samples of leakers and non-leakers is small ($\Delta \beta \lesssim 0.2$). The lack of significant variations in $\beta$ possibly indicates that the indirect tracers of LyC based on the \lya{} line profile are not as sensitive to the dust content as the $\beta$ slope.

\section{Spectral stacking}
\label{sec:stacks}
    
    \subsection{Method}
    Most of the rest-frame UV lines, both in emission and absorption, are too faint to be detected in the individual spectra of our LAEs. To address this problem, we perform spectral stacking to increase the S/N of the observed spectra.
    
    First, we convert wavelengths from air into vacuum conditions\footnote{\url{http://www.astro.uu.se/valdwiki/Air-to-vacuum\%20conversion}} and shift the spectra to the rest-frame. We use systemic redshifts recovered from the \lya{} line profiles following the method described in \citet[hereafter V18]{Verhamme2018}. They propose two diagnostics derived from the spectroscopic observations of LAEs with accurate systemic redshift measurements. In the case of the low- and high-$\upsilon_\text{sep}$ subsamples, we use Eq.~(1) of V18 which relates $\upsilon^{\text{red}}_{\text{peak}}$ to the \lya{} peak separation. For other subsamples, we use the empirical correlation between $\upsilon^{\text{red}}_{\text{peak}}$ and the \lya{} FWHM (Eq.~(2) of V18). We discuss possible caveats associated with this approach in \autoref{sec:systemic_redshift}.
    
    The rest-frame spectra of galaxies can have a varying wavelength sampling depending on their redshift. Therefore, we resample the spectra using the flux-conserving \textsc{SpectRes} tool \citep{Carnall2017}. We adopt a target sampling of 0.25~\r{A} corresponding to a rest-frame sampling of a spectrum observed by MUSE at $z = 4$. We then create the median-stacked spectra and estimate uncertainties of the spectral flux densities using the standard deviation of 200~bootstrap replications. We apply the median statistics instead of the mean or the weighted average to ensure that the stacked spectra are not dominated by the few brightest sources in our sample.
    
    \begin{figure*}
        \includegraphics[width=\linewidth]{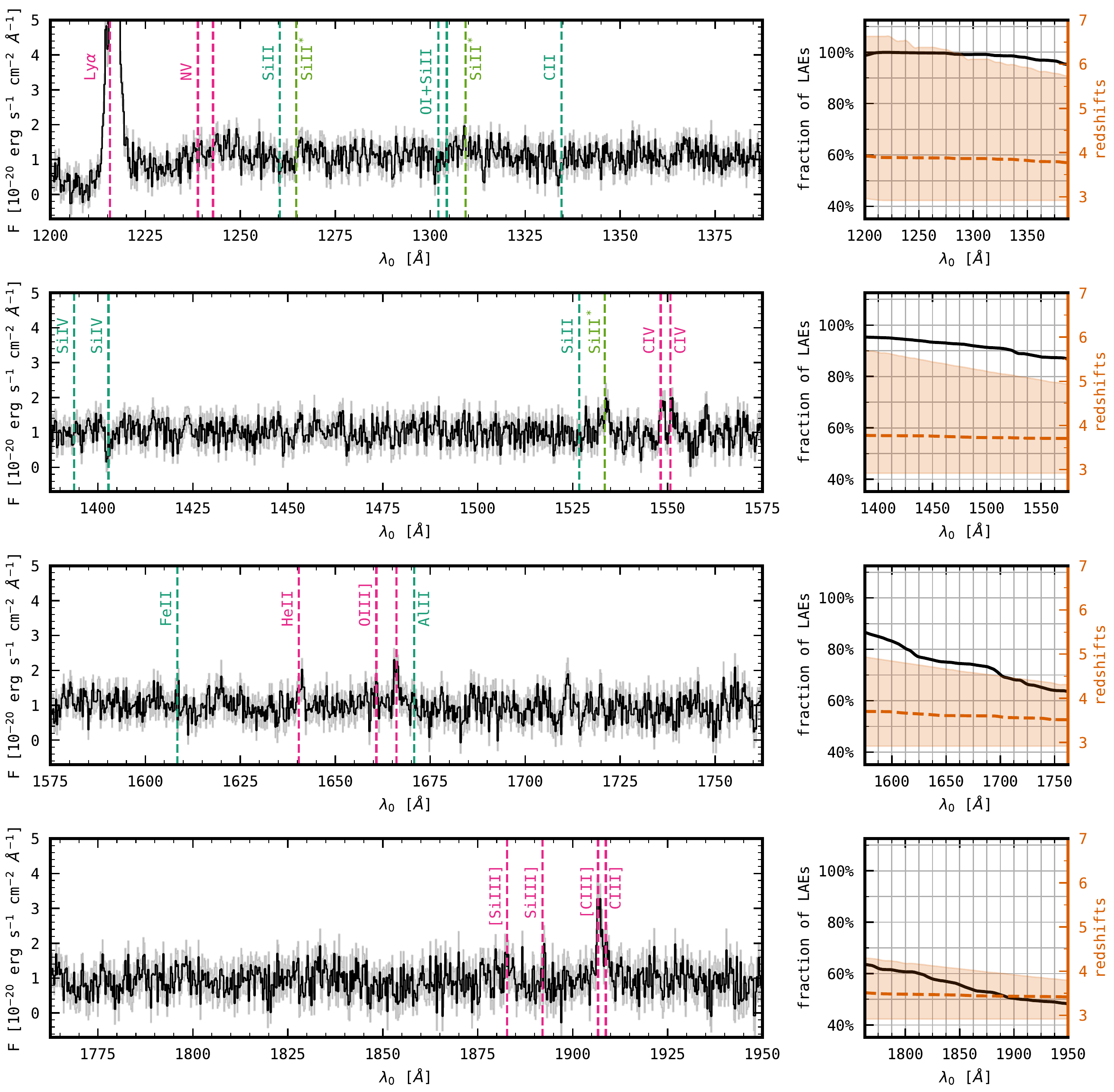}
        \caption{\textbf{Left:} Median-stacked spectra of the total sample of MUSE LAEs described in \autoref{sec:data}. Grey shaded regions represent the 1$\sigma$ noise level. The stacked spectra reveal various nebular emission lines (pink), ISM absorption lines (green) and fine-structure transitions (yellow-green) typically observed in high-redshift and local metal-poor star-forming galaxies. Dashed vertical lines mark the rest-frame wavelengths of the spectral lines. \textbf{Right:} Fraction of LAEs included in the stack (black solid line) and redshift coverage (light orange area) as a function of wavelength. The median redshift is shown by the dashed orange line.}
        \label{fig:full_stack}
    \end{figure*}
    
    \subsection{Rest-frame UV lines revealed by the stacked spectra}
    The median-stacked spectra for the total LAE sample are shown in \autoref{fig:full_stack}. We apply a median filter with a window size of 100~\r{A} to fit the continuum and obtain an average S/N of $3.1$ per spectral bin (0.25~\r{A}) in the wavelength range $1250\;\text{\AA} < \lambda < 1900\;\text{\AA}$. This value significantly exceeds a typical S/N of an individual LAE spectrum ($\lesssim 0.1$), demonstrating the efficiency of spectral stacking.
    
    In the same figure, we mark the rest-frame wavelengths of various nebular emission lines, ISM absorption lines and fine-structure transitions which are commonly observed in the spectra of high-redshift and local metal-poor star-forming galaxies (see \citealp{Feltre2020} and references therein). We measure the line EWs by fitting Gaussian functions to the continuum-subtracted spectra sliced to $\pm 500$~\kms{} regions around the lines. For the fitting procedure, we use the \textsc{astropy} implementation of the the Levenberg-Marquardt optimisation algorithm, constraining the FWHM within the range of $50-300$~\kms{}. If the fit is unsuccessful, we measure the EW non-parametrically by summing up the flux in the same spectral region where the fit is performed. In the case of the \ion{[C}{III]}$\lambda$1907$+$\ion{C}{III]}$\lambda$1909 doublet (hereafter \ion{C}{III]}), we use a sum of two Gaussian profiles with a tied peak separation of 2.05~\r{A}. We fit Gaussian functions to the rest of the line doublets separately because their components do not overlap with each other. In the case of the ISM absorption lines, we shift the spectral regions by $-200$~\kms{} to take into account absorption line velocity offsets due to large-scale gas outflows \citep[e.g.][]{Shapley2003}. Finally, we estimate the EW uncertainties using a Monte-Carlo approach by repeating the measurements on the spectra perturbed with noise 1000 times.
    
    Among the high-ionisation emission lines, we report a $\text{S/N} > 3$ detection of the \ion{C}{IV}$\lambda\lambda$1548,1551 resonant doublet (hereafter \ion{C}{IV}), collisionally excited \ion{O}{III]}$\lambda\lambda$1661,1666 and \ion{C}{III]} doublets, and the \ion{He}{II}$\lambda1640$ line (hereafter \ion{He}{II}; the EWs are reported in \autoref{tab:line_ews}). We measure the EW of only the \ion{O}{III]}$\lambda$1666 component of the doublet (hereafter \ion{O}{III]}), because the weaker bluer component (\ion{O}{III]}$\lambda$1661) is not detected in our stacks with $\text{S/N} > 3$. Additionally, we find \ion{[Si}{III]}$\lambda$1883+\ion{Si}{III]}$\lambda$1892 nebular emission in some of the stacks (\autoref{sec:results}). We limit ourselves to the analysis of the \ion{[Si}{III]}$\lambda$1883 component (hereafter \ion{[Si}{III]}) because it always dominates the total flux of the doublet.
    
    While \ion{O}{III]}, \ion{C}{III]} and \ion{[Si}{III]} represent exclusively nebular emission, the origins of \ion{He}{II} are likely more complex. In particular, strong and dense stellar winds of Wolf$-$Rayet (W$-$R) stars give rise to broad \ion{He}{II} features observed along with nebular \ion{He}{II} \citep[e.g.][]{Nanayakkara2019}. The \ion{C}{IV} line also includes other components in addition to nebular emission, i.e. emission from stellar winds in O and B stars characterised by P~Cygni profiles, and ISM absorption \citep[e.g.][]{Berg2018}. The S/N of our stacked spectra is insufficient to study the \ion{He}{II} and \ion{C}{IV} line profiles in detail in order to constrain contributions from different sources of emission (and absorption). Therefore, we report only the total flux of \ion{He}{II} and \ion{C}{IV} obtained from fitting Gaussian profiles.
    
    Finally, we compare our stacked spectra with the full stack of 220 LAEs from the MUSE HUDF survey shown in Figure~3 of \citet{Feltre2020}. Their spectra reveal a similar collection of nebular emission lines (e.g. \ion{O}{iii]}, \ion{He}{ii} and \ion{C}{iii]}) and ISM absorption features (e.g. \ion{C}{ii}$\lambda$1335 and \ion{Si}{iv}$\lambda$1403). However, they obtain a higher S/N per spectral pixel ($\text{S/N}\sim5$ vs. $\text{S/N}\sim3$) despite using a similar wavelength sampling (0.3~\r{A} vs. 0.25~\r{A}), likely because their sample only includes the deep MUSE HUDF data. On the other hand, by including the MUSE-Wide LAEs in our sample, we are probing both bright \textit{and} faint LAEs, thus expanding the parameter space. This enables a more comprehensive analysis of the galaxy properties related to LyC escape.
    
    \subsection{The impact of redshift uncertainties on measuring EWs}
    \label{sec:systemic_redshift}
\begin{figure}
        \includegraphics[width=\linewidth]{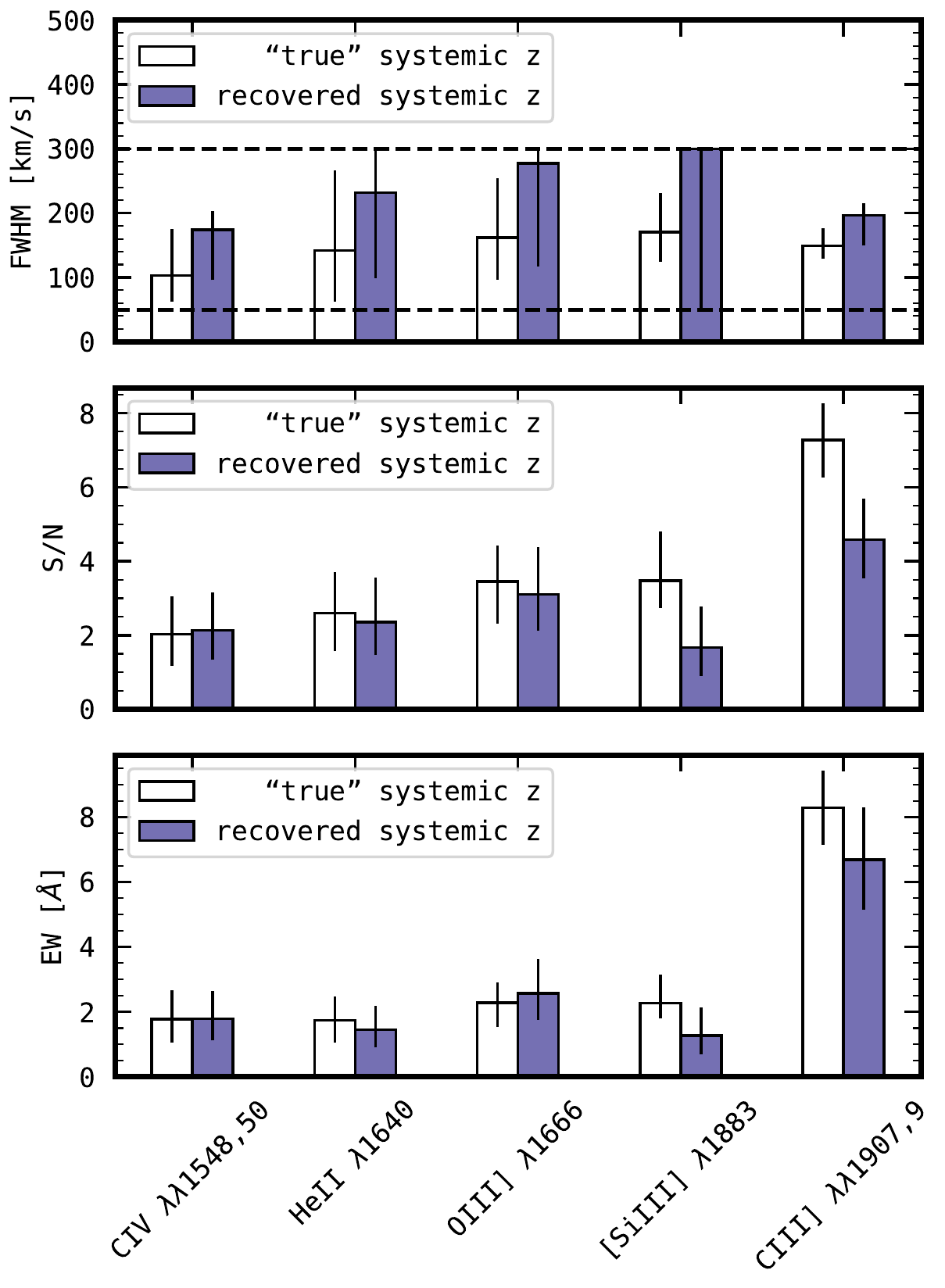}
        \caption{Comparison between the properties of the rest-frame UV emission lines detected in the median-stacked spectra of 24~LAEs with at least one non-resonant line detection at $\text{S/N} > 3$. Before stacking, the individual spectra are shifted to the rest-frame using systemic redshifts estimated from (i) velocity offsets of non-resonant lines (white), and (ii) the empirical relation between the \lya{} velocity offset and the \lya{} FWHM (Eq.~(2) of V18; dark blue). \textbf{Top to Bottom:} Line FWHMs, $\text{S/N}$ ratios and EWs. Vertical error bars mark the 16-th and the 84-th percentiles of the distributions of the measured quantities. Dashed horizontal lines in the top panel indicate the bounded constraints applied to the line FWHMs in the fitting procedure. The empirical method used to estimate systemic redshifts results, on average, in line broadening, a decrease in the S/N and EW underestimation.}
        \label{fig:systemic_z}
    \end{figure}

    Empirical relations used to recover systemic redshifts provide lower accuracy than the direct estimates based on the shifts of non-resonant lines. If systemic redshifts are measured with high statistical uncertainties, line fluxes are spread over a large wavelength range in the spectral stacks, resulting in a low S/N and making the assessment of the line properties more difficult. If the spectra are stacked using the median statistics, the line EWs could in addition be biased towards lower values. For example, \citet{Feltre2020} stacked copies of idealised spectra with a line center shifted according to Eq.~(2) of V18 and found that the EW is underestimated by $15-20$~per~cent. To investigate these effects, we select 24~LAEs with detected ($\text{S/N} > 3$) non-resonant emission lines, i.e. collisionally excited nebular emission lines and the \ion{He}{II} line. We compare two spectral stacks obtained using (i) the ``true'' systemic redshifts based on the non-resonant line velocity offsets with respect to the line rest wavelengths, and (ii) systemic redshift estimates given by Eq.~(2) of V18. We find that if systemic redshifts are recovered using the \lya{} line properties (Eq.~(2) of V18), the EW measurements have similar absolute uncertainties but the line profiles are 1.2--1.6~times broader (\autoref{fig:systemic_z}, top panel) and the line S/N-s are typically lower by a few tens of per cent (\autoref{fig:systemic_z}, middle panel). The bottom panel in \autoref{fig:systemic_z} shows that the EWs of \ion{C}{IV}, \ion{He}{II} and \ion{C}{III]} are underestimated by 5--20~per~cent, in agreement with the simulation results from \citet{Feltre2020}. However, we find that the \ion{[Si}{III]} EW is underestimated by as much as 44$^{+39}_{-27}$~per~cent, whilst the \ion{O}{III]} EW is higher than the ``true'' EW by 15$^{+78}_{-42}$~per~cent. Taking into account high statistical uncertainties of these measurements, larger samples of LAEs with known systemic redshifts are required to provide a more accurate estimate of the EW bias. Nevertheless, we caution that the EWs reported in this paper might in some cases be underestimated by up to $\approx20$~per~cent.

    \subsection{Smaller redshift coverage at longer wavelengths}
    \label{sec:redshift_coverage}
    The effective number of objects used to compute the median stack varies as a function of wavelength since our LAEs are observed at different redshifts (right panels in \autoref{fig:full_stack}). Starting from 1422 sources (100~per~cent) at the \lya{} rest-frame wavelength (1215.67~\r{A}), this number gradually decreases towards longer wavelengths, reaching 721 (51~per~cent) at 1900~\r{A} (at $\lambda > 1900$~\r{A} the S/N quickly drops down). At the same time, the maximum possible redshift of an object included in the stack decreases from $z_{\text{max}}=6.7$ at 1215~\r{A} to $z_{\text{max}}=3.9$ at 1900~\r{A}. Despite having smaller redshift coverage at longer wavelengths, we can expect that e.g. the properties of the \ion{C}{III]} line observed in LAEs at $2.9 < z < 3.9$ would be representative for LAEs at $4 \lesssim z \lesssim 6$ as well. This assumption is based on the observational evidence that LAEs at $2 \lesssim z \lesssim 6$ share a similar distribution of several fundamental properties including sizes \citep[e.g.][]{Malhotra2012}, UV slopes, \lya{} EWs and scale length parameters \citep[e.g.][]{Santos2020}, and \lya{} line profiles corrected for the IGM absorption \citep{Hayes2021}. In addition, the observed LAE luminosity function does not evolve significantly at such redshifts \citep[e.g.][]{Cassata2011, Herenz2019}. We also compare the full stack and the stack of LAEs at $2.9 < z < 3.9$, and find that the $\ion{C}{IV}/\ion{C}{III]}$ ratio differs by only $\sim6$~per~cent. Finally, we note that the \textit{median} redshift does not change significantly over the same wavelength range, decreasing from $z_{\text{median}}=3.9$ at 1215~\r{A} to $z_{\text{median}}=3.4$ at 1900~\r{A}. Throughout the rest of this paper, we assume that the evolution of redshift coverage with wavelength has minimal impact on the observed line properties.

\section{Ionising properties and physical conditions in the ISM of LyC-leaker candidates}
\label{sec:results}

\begin{figure*}
        \includegraphics[width=\linewidth]{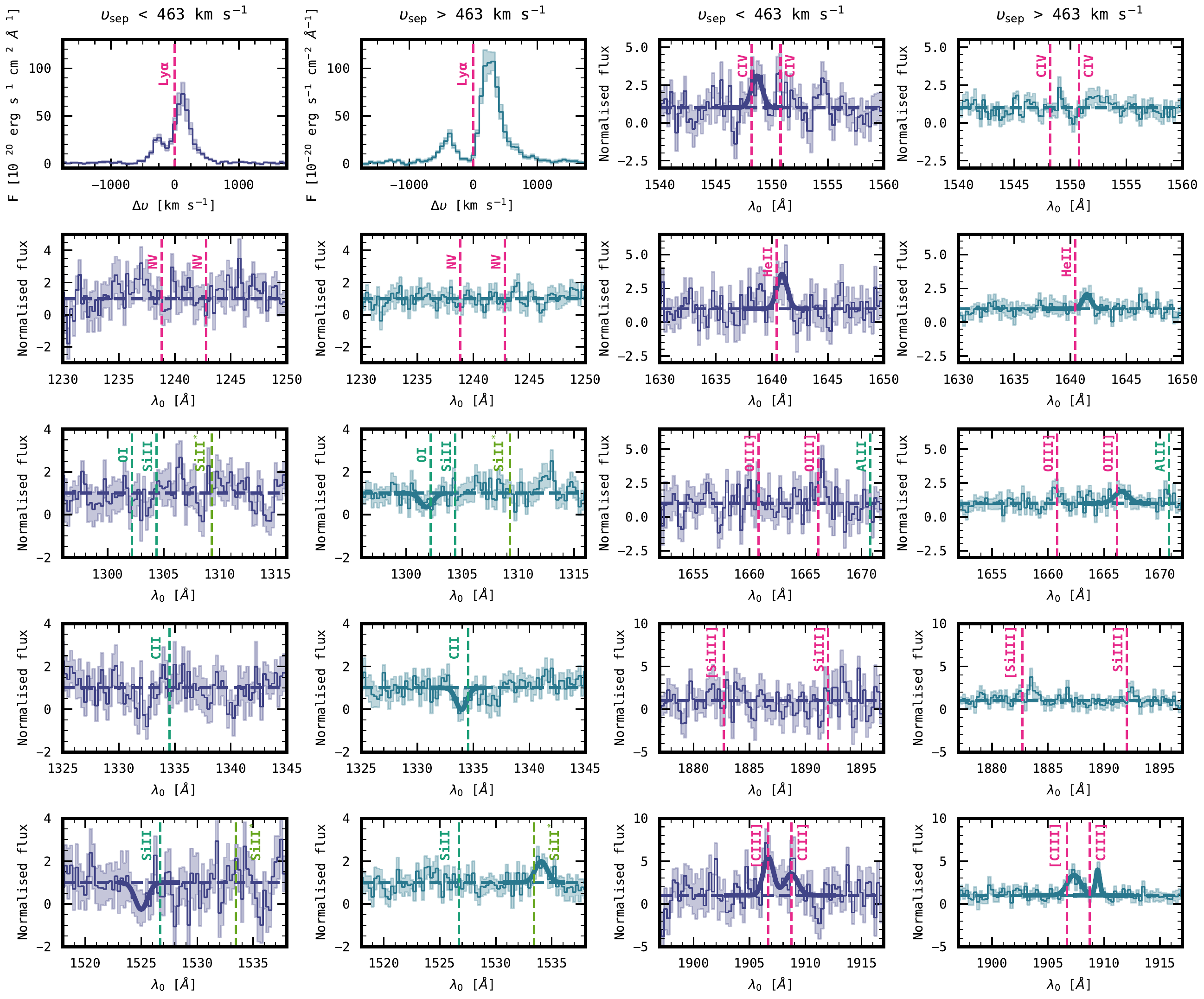}
        \caption{Median-stacked spectra of LAEs with low ($<463$\kms{}; blue) and high ($>463$\kms{}; green) \lya{} peak separations. Each panel represents a spectral region around one of the rest-frame UV lines (or groups of lines) whose rest-frame wavelengths are indicated by vertical pink (nebular emission), green (ISM absorption) and yellow-green (fine-structure transition) dashed lines. Shaded regions represent the 1$\sigma$ noise level computed via the bootstrap method. Horizontal dashed lines mark the continuum level. Gaussian profiles fitted to the spectral lines are shown for $\text{S/N}>3$ detections (thick solid lines).}
        \label{fig:peaksep_stacks}
    \end{figure*}

    In this section, we investigate the rest-frame UV spectral properties of our LyC-leaker candidates selected as having narrow, symmetric \lya{} profiles with small peak separation (\autoref{sec:subsamples}). We apply the stacking technique described in \autoref{sec:stacks} to each group of galaxies. In \autoref{fig:peaksep_stacks}, we show the median-stacked spectra for both low- and high-$\upsilon_{\text{sep}}$ subsamples, with a focus on the spectral regions around the rest-frame UV lines. We also show the Gaussian profiles fitted to the spectral lines with a $\text{S/N} > 3$ detection. Similar figures but for the groups of different \lya{} FWHMs and $a_{\text{asym}}$ are presented in the appendix (\autoref{appendix:stacks}).

\begin{figure*}
        \includegraphics[width=\linewidth]{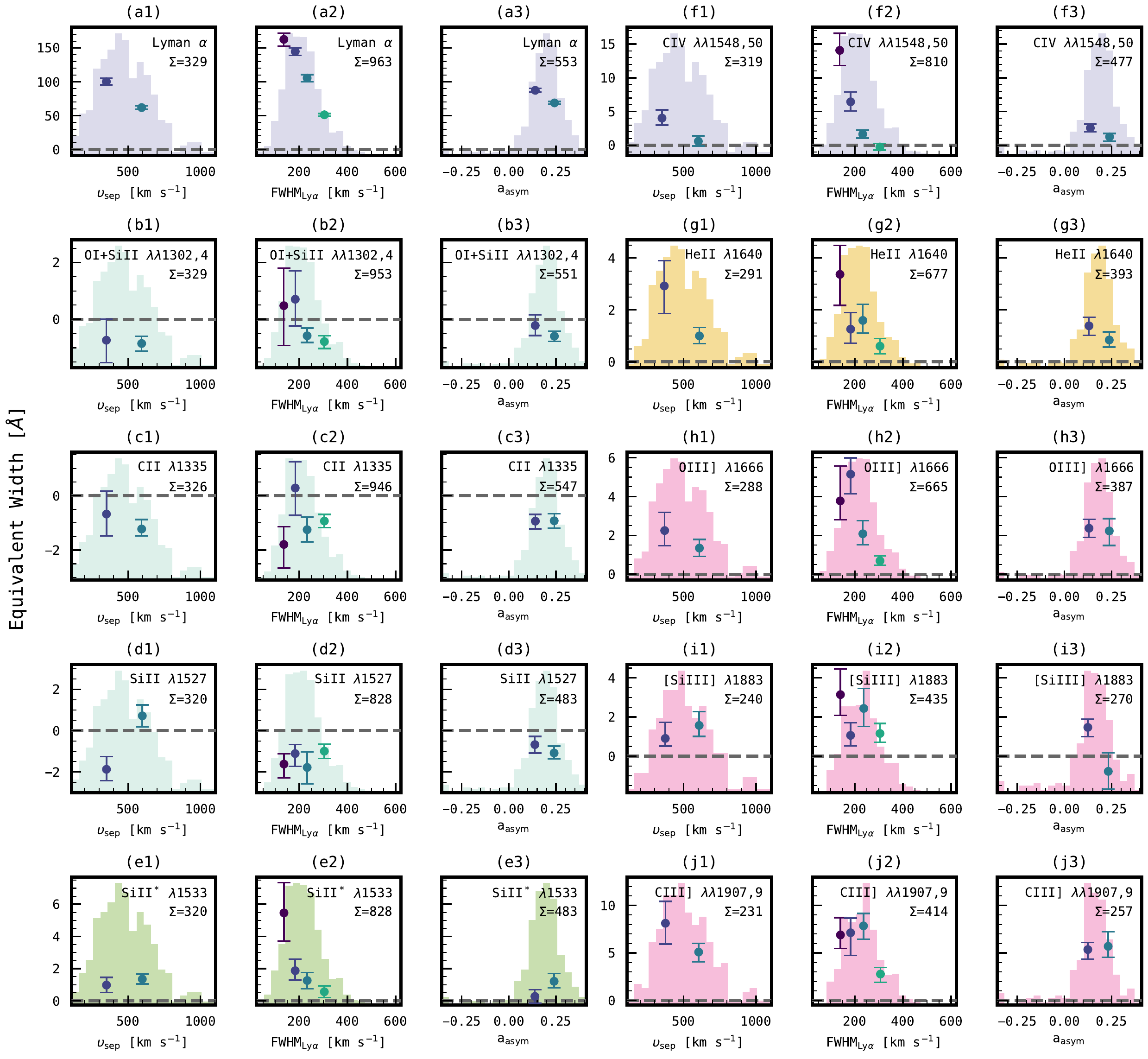}
        \caption{EWs of rest-frame UV lines detected in the stacked spectra of candidates for LyC-leakers and non-leakers. Markers show the EWs measured for each group of LAEs. Histograms show the distributions of the \lya{} line properties. \textbf{Left column:} \lya{} (lavender; panels a1-a3), ISM absorption lines (sea foam green; panels b1-d3) and the \ion{Si}{II}$^{*}\lambda$1533 fine-structure transition (moss green; panels e1-e3). \textbf{Right column:} \ion{C}{IV} (lavender; panels f1-f3), \ion{He}{II} (yellow; panels g1-g3) and collisionally excited nebular emission lines (pink; panels h1-j3). The color-coding of markers follows that of the stacked spectra of respective subsamples in \autoref{fig:peaksep_stacks}, \autoref{fig:fwhm_stacks} and \autoref{fig:asym_stacks}. The distributions of the \lya{} line properties are comprised of objects included in the stacks at respective wavelengths; the number of objects is shown in the top right corners of panels.}
        \label{fig:line_ews}
    \end{figure*}

    The EWs of nebular emission lines, LIS absorption lines and fine-structure emission lines detected with $\text{S/N} > 3$ in at least one of the stacks are listed in \autoref{tab:line_ews} and plotted in \autoref{fig:line_ews}. Apart from \lya{}, one of the strongest emission lines we observe is \ion{C}{III]}, with the total EW ranging from $\sim$~3~\r{A} (FWHM$_{\text{Ly}\alpha}$, Q4) to $\sim$~8~\r{A} (the low-$\upsilon_{\mathrm{sep}}$ subsample; see column 12 in \autoref{tab:line_ews}). We resolve the double-peaked profile of \ion{C}{III]} in all the stacks except for FWHM$_{\text{Ly}\alpha}$, Q1, for which we fit a single peak.
    Next, we report a $\text{S/N} > 3$ detection of the \ion{C}{IV} line in the low-$\upsilon_{\mathrm{sep}}$ and -$a_{\mathrm{asym}}$ groups, and the two lowest FWHM$_{\text{Ly}\alpha}$ quartiles (Q1 and Q2), with the EW as high as $14 \pm 2$~\r{A} (FWHM$_{\text{Ly}\alpha}$, Q1; see column 8 in \autoref{tab:line_ews}). We do not find significant ($>2\sigma$) \ion{C}{IV} absorption in any of the stacks, suggesting that the contribution from the stellar winds of OB stars to \ion{C}{IV} is negligible compared to the nebular emission from ionised gas. Finally, the EWs of \ion{He}{II}, \ion{O}{III]} and \ion{[Si}{III]} take values between $\sim$~1~\r{A} and $\sim$~4~\r{A}, with an exception of \ion{[Si}{III]} in the high-$a_{\mathrm{asym}}$ subsample in which the EW is consistent with zero (see columns 9-11 in \autoref{tab:line_ews}). \ion{[Si}{III]} is detected with $\text{S/N} > 3$ only in the low-$a_{\mathrm{asym}}$ stack, whereas \ion{He}{II} -- also in the two $\upsilon_{\mathrm{sep}}$ stacks, and \ion{O}{III]} -- in both the low- and high-$a_{\mathrm{asym}}$ stacks, the high-$\upsilon_{\mathrm{sep}}$ stack and the two intermediate FWHM$_{\text{Ly}\alpha}$ quartiles.

    By measuring the line EWs and the line ratios, we qualitatively compare the properties of the ionising sources (\autoref{sec:ionising_sources} -- \autoref{sec:spectral_hardness}) and the physical conditions in the ISM (\autoref{sec:ism_geometry} and \autoref{sec:ism_opacity}) of the LyC-leaker and non-leaker candidates. We discuss the relationships between the production of ionising photons, nebular emission, ISM absorption, and \lya{} and LyC escape.
    
    \subsection{Overture: source(s) of the ionising radiation}
    \label{sec:ionising_sources}
\begin{figure}
        \includegraphics[width=\linewidth]{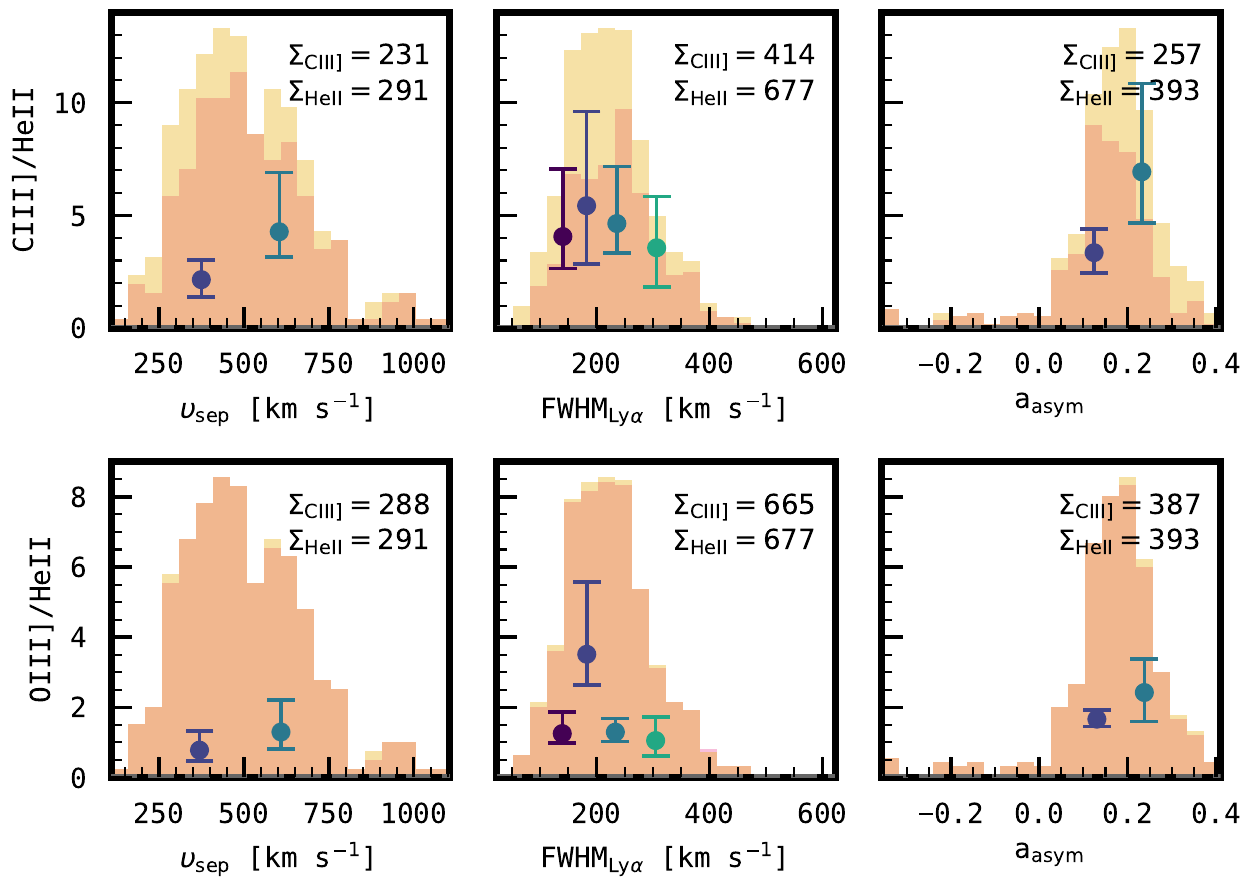}
        \caption{\ion{[C}{III]}$\lambda$1907$+$\ion{C}{III]}$\lambda$1909$/$\ion{He}{II}$\lambda$1640 and \ion{O}{III]}$\lambda$1666$/$\ion{He}{II}$\lambda$1640 flux ratios for the LAE subsamples described in \autoref{sec:subsamples}. The observed line ratios are consistent with the photoionisation models of star-forming galaxies \citep[e.g.][]{Feltre2016}, which rules out a significant contribution of AGNs to the nebular emission detected in the stacks. Background histograms are as in \autoref{fig:line_ews}.}
        \label{fig:line_ratios}
    \end{figure}

    Determining the source(s) of ionising photons is one of the main goals in reionisation studies. We argue that AGNs cannot dominate the ionising radiation from our LAEs. First, we carry out the spectroscopic diagnostics based on the line ratios which trace the presence of an AGN. We find that $\log{(\ion{C}{III]} / \ion{He}{II})} > 0$ and $\log{(\ion{O}{III]} / \ion{He}{II})} > -0.5$ (\autoref{fig:line_ratios}), in contradiction with the pure AGN photoionisation models \citep[][Fig.~A1]{Feltre2016}. Second, we detect moderate strengths of \ion{C}{III]} ($\text{EW} \sim 3-8$~\r{A}; see \autoref{tab:line_ews}, column 12) and \ion{C}{IV} ($\text{EW} \sim 0-14$~\r{A}; ibid., column 8), which are not compatible with the AGN scenario either \citep[$\text{EW} > 20$~\r{A} and $\text{EW} > 12$~\r{A}, respectively;][]{Nakajima2018}. Furthermore, we observe a lack of \ion{N}{V} emission (second row in \autoref{fig:peaksep_stacks}, \autoref{fig:fwhm_stacks} and \autoref{fig:asym_stacks}) -- a sign of the AGN activity due to a very high photon energy required for ionisation of this element \citep[$>77.4$~eV; e.g.][]{Sobral2018}. Thus, we suggest that young, metal-poor stars have a dominant contribution to the ionising photon budget of our LAEs, with no exception for the LyC-leaker candidates.
    
    However, additional ionising sources might be needed to explain the \ion{He}{II} EWs ($\sim 1-3$~\r{A}; see \autoref{tab:line_ews}, column 9). Although we cannot disentangle the nebular and stellar components of \ion{He}{II} to examine its possible origins, we note that the current photoionisation models are unable to fully reproduce the observed \ion{He}{II} emission \citep[e.g.][]{Nanayakkara2019}. Various mechanisms accounting for the missing ionising photons are proposed to resolve this tension, including emission from Pop~III stars, radiative shocks, increased stellar rotation, production of stripped stars and emission from X-ray binaries and ultra-luminous X-ray sources \citep[e.g.][]{Izotov2012,Eldridge2012,Smith2018,Schaerer2019,Simmonds2021}.
    
    \subsection{Nebular emission lines: the ionising photon budget}
    \label{sec:ionising_budget}
    In the framework of the density-bounded model of the ISM, high escape fractions of LyC are achieved in a nearly completely ionised medium \citep[e.g.][]{Nakajima2014}. Therefore, efficient LyC escape suggests high ionising fluxes from the regions of intense star formation. To test this hypothesis, we inspect the nebular emission lines, whose properties provide insight into the ionising photon budget of a galaxy. More specifically, strong nebular emission lines indicate low, sub-solar gas-phase metallicities and high ionisation parameters \citep[e.g.][]{Steidel2016,Jaskot2016,Senchyna2017} which imply the presence of young stellar populations producing copious amounts of ionising photons. We find higher EWs of nebular \ion{O}{III]}, \ion{[Si}{III]} and \ion{C}{III]} predominantly in the low-$\upsilon_{\mathrm{sep}}$ and -FWHM$_{\text{Ly}\alpha}$ stacks (\autoref{fig:line_ews}, panels h1-h2, i1-i2 and j1-j2), supporting the idea that LyC leakers have an increased production rate of ionising photons compared to non-leakers \citep[see also][]{Naidu2022}. The EWs of \lya{}, \ion{C}{iv} and \ion{He}{ii} show even more compelling trend (ibid., panels a1-a2, f1-f2 and g1-g2). However, we caution that (i) \ion{He}{ii} might have a significant stellar component in addition to the nebular (see \autoref{sec:ionising_sources}), (ii) \lya{} and \ion{C}{iv} are resonant lines, a priori sensitive to the radiation transfer effects which could modulate their EWs alongside with LyC~$f_{\text{esc}}$ (see \autoref{sec:ism_opacity}).
    
    The LAE-selected sample gives us the opportunity to use the \lya{} line itself to select galaxies with substantially different production rates of ionising photons. We create four \lya{} EW and luminosity quartiles using the data from \citet{Kerutt2022}. We find that the EWs of \textit{all} the nebular emission lines increase dramatically with the \lya{} EW (\autoref{fig:line_ews_app}, third column), in good agreement with previous studies \citep[e.g.][]{Shapley2003,Stark2014,Feltre2020}. This correlation strongly suggests that high-redshift galaxies with higher \lya{} EWs have higher ionising photon production efficiency ($\xi_{\text{ion}}$), implying stellar populations with lower metallicities and younger ages as demonstrated by \citet{Maseda2020} for a sample of continuum-faint LAEs at $z \approx 4-5$. The stacks of LAEs with different \lya{} luminosities, on the other hand, have nebular emission lines of similar strengths (\autoref{fig:line_ews_app}, fourth column), which could be explained by a variety of metallicites and ages presented at fixed \lya{} luminosity.
    
    \subsection[He II: hardness of the ionising spectrum]{\ion{He}{ii}: hardness of the ionising spectrum}
    \label{sec:spectral_hardness}
    Whether or not a hard ionising spectrum is a necessary condition for efficient LyC escape remains a subject of debate \citep{Naidu2022,Schaerer2022,Marques-Chaves2022,Enders2023}. We search for evidence of a harder ionising spectrum in the stacks of LyC-leaker candidates by inspecting \ion{He}{II} emission. We find that the \ion{He}{II} EW is typically a few times higher among potential LyC leakers (\autoref{fig:line_ews}, panels g1-g3), suggesting an elevated production rate of \ion{He}{}$^+$-ionising photons with energies $>54.4$~eV. In addition, we report lower $\ion{C}{iii]}/\ion{He}{ii}$ and $\ion{O}{iii]}/\ion{He}{ii}$ ratios in the low-$\upsilon_{\mathrm{sep}}$ and -$a_{\mathrm{asym}}$ stacks  (\autoref{fig:line_ratios}) which possibly indicates a harder ionising spectrum, given that the ionisation energies for \ion{C}{iii]} and \ion{O}{iii]} are much lower compared to that of \ion{He}{II} (24.4~eV and 35.1~eV, respectively). Similar results were obtained by \citet{Naidu2022} who studied composite spectra of 26 LAEs at $z \sim 2$ from the X-SHOOTER \lya{} survey (XLS-$z$2). They detected prominent narrow \ion{He}{II} (and \ion{C}{IV}) emission in the ``high escape'' stack (i.e. galaxies with low $\upsilon_{\mathrm{sep}}$ \textit{or} high fraction of the \lya{} flux at nearly systemic velocity), and only lower-ionisation lines (e.g. \ion{C}{iii]} and \ion{O}{iii]}) in the ``low escape'' stack (i.e. galaxies with high $\upsilon_{\mathrm{sep}}$ \textit{and} low fraction of the \lya{} flux at nearly systemic velocity).
    
    At low redshift ($z \sim 0.3-0.4$), \citet{Schaerer2022} studied rest-frame UV spectra of eight LyC emitters and found that the galaxies with $f_{\text{esc}} > 10$~per~cent show strong \ion{He}{II} emission with the rest-frame EWs ranging from 3 to 8~\r{A}. They argued that such high values are primarily due to elevated $\xi_{\text{ion}}$, and that galaxies with high $f_{\text{esc}}$ do not exhibit exceptionally hard ionising spectra compared to other galaxies at similar metallicities. Recently, \citet{Marques-Chaves2022} studied a large sample of SFGs from the HST Low-$z$ Lyman Continuum Survey and found no correlation between \ion{He}{II}$\lambda$4686$/$H$\beta \lambda$4861 and $f_{\text{esc}}$, arriving at the same conclusions. Our results suggest that LyC leakers indeed have higher $\xi_{\text{ion}}$ (\autoref{sec:ionising_budget}), however we cannot rule out the possibility that higher $f_{\text{esc}}$ also accompanies harder radiation fields, especially in the light of our $\ion{C}{iii]}/\ion{He}{ii}$ and $\ion{O}{iii]}/\ion{He}{ii}$ measurements.
    
    \begin{figure*}
        \includegraphics[width=\linewidth]{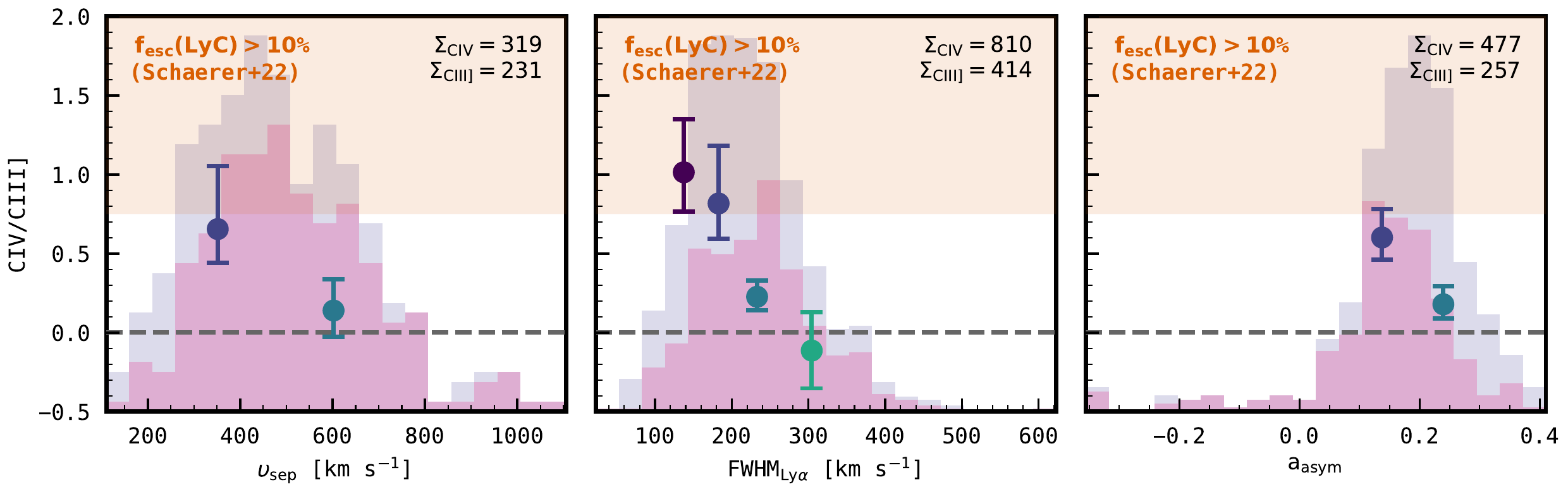}
        \caption{\ion{C}{IV}$\lambda\lambda$1548,1550 to \ion{[C}{III]}$\lambda$1907$+$\ion{C}{III]}$\lambda$1909 flux ratio for the LAE subsamples described in \autoref{sec:subsamples}. The observed line ratios measured for the LyC-leaker candidates (purple and dark blue) are consistent with \ion{C}{IV}/\ion{C}{III]} ratios measured by \citet{Schaerer2022} for a sample of low-redshift ($z \sim 0.3-0.4$) galaxies with $f_{\text{esc}} > 10$\% ($\ion{C}{IV}/\ion{C}{III]} \gtrsim 0.75$; shaded area). The rapid growth of \ion{C}{IV}/\ion{C}{III]} with increasing $f_{\text{esc}}$ suggests that this line ratio could serve as an alternative indirect tracer of $f_{\text{esc}}$ at redshifts where \lya{} is significantly attenuated by the IGM ($z \gtrsim 6$). Background histograms are as in \autoref{fig:line_ews}.}
        \label{fig:civ_ciii}
    \end{figure*}
    
    \subsection[LIS absorption lines and Si II*: the ISM geometry]{LIS absorption lines and \ion{Si}{ii}$^{*}$: the ISM geometry}
    \label{sec:ism_geometry}
    Interestingly, the EWs of \ion{O}{III]}, \ion{C}{III]} and even \ion{He}{II} (but not \lya{}, \ion{C}{iv} and \ion{[Si}{III]}) remain constant within 1~$\sigma$ uncertainties if we select LyC-leaker candidates using $a_{\text{asym}}$ (\autoref{fig:line_ews}, panels h3, j3 and g3). We suggest that the specifics of the ISM geometry could break the link between nebular emission and $f_{\text{esc}}$. Unlike $\upsilon_{\mathrm{sep}}$ and FWHM$_{\text{Ly}\alpha}$ which likely probe the density-bounded geometry, the asymmetry of the \lya{} line profile might be more sensitive to the ``picket-fence'' geometry, i.e. a clumpy ISM with open lines of sight (LOS). In this scenario, high $f_{\text{esc}}$ does not require constant support by a strong ionising radiation field because LyC photons escape freely through empty channels in the ISM.
    
    For the ``picket-fence'' geometry, a canonical parameter measuring the fraction of LOS covered by dense clumps of neutral hydrogen which block LyC radiation is $f_{\text{cov}}$, or the neutral gas covering fraction. Low $f_{\text{cov}}$ (or, equivalently, high $f_{\text{esc}}$) likely manifests itself in residual fluxes in the cores of LIS absorption lines \citep[e.g.][]{Heckman2011,Saldana-Lopez2022}. In agreement with such interpretation of absorption line measurements, we find that the EWs of \ion{O}{I}$\lambda$1302$+$\ion{Si}{II}$\lambda$1304 and \ion{Si}{II}$\lambda$1527 are lower for the stacks with more symmetric \lya{} profiles (\autoref{fig:line_ews}, panels b3 and d3). The EW of \ion{C}{II}$\lambda$1335 is about $-0.9$~\r{A} for both stacks (\autoref{fig:line_ews}, panel c3), but some additional flux bluer to the fitted Gaussian is probably lost in the high-$a_{\text{asym}}$ stack (see \ref{fig:asym_stacks}). Conversely, the EWs of the LIS absorption lines detected in the $\upsilon_{\mathrm{sep}}$ and FWHM$_{\text{Ly}\alpha}$ stacks do not show any clear trend (\autoref{fig:line_ews}, panels b1-b2, c1-c2, d1-d2), suggesting that $f_{\text{cov}}$ is loosely coupled to $\upsilon_{\mathrm{sep}}$ and FWHM$_{\text{Ly}\alpha}$.
    
    We caution that the LIS absorption line depths are also affected by the metallicity and the kinematics of the gas, and the infilling by resonant emission \citep[e.g.][]{Vasei2016}. Moreover, the LIS diagnostics is hampered by the low S/N data. Such effects could explain the lack of the correlation between the LIS absorption line EWs and the \lya{} EW (\autoref{fig:line_ews_app}, panels b1-d1) which was reported by previous studies \citep[e.g.][]{Shapley2003,Jones2012}. This result suggests that a higher sensitivity is required for more rigorous analyses of the LIS absorption lines.
    
   Both the density-bounded and ``picket-fence'' geometries represent idealised physical models of the ISM. To bring them into a unified picture, we examine the overall distribution of neutral gas in our galaxies. The neutral material off the LOS can be probed with ﬂuorescent non-resonant emission lines such as \ion{Si}{ii}$^{*}\lambda$1533 \citetext{hereafter \ion{Si}{ii}$^{*}$; e.g. \citealp{Jaskot2014}}. We detect \ion{Si}{ii}$^{*}$ emission at a significance level of $>2\sigma$ in \textit{all} of the $\upsilon_{\mathrm{sep}}$ and FWHM$_{\text{Ly}\alpha}$ stacks, with the highest \ion{Si}{ii}$^{*}$ EW in the lowest FWHM$_{\text{Ly}\alpha}$ quartile (\autoref{fig:line_ews}, panels e1-e2). Strong \ion{Si}{ii}$^{*}$ in the stacks of LyC-leaker candidates suggests large amounts of \ion{H}{I} off the LOS, contradicting the ``classical'' density-bounded scenario which assumes a largely isotropic, almost completely ionised ISM. \lya{} (and LyC) photon escape in an anisotropic medium could be attributed to the presence of highly ionised channels of low optical depth \citep[e.g.][]{Zackrisson2013,Rivera-Thorsen2017}. In this case, the relative transparency of these channels determines $\upsilon_{\mathrm{sep}}$ and FWHM$_{\text{Ly}\alpha}$ \citep{McKinney2019,Jaskot2019,Kakiichi2021}. The distribution of high-column-density gas, in turn, might be more relevant to $f_{\text{cov}}$ in the context of the ``picket-fence'' geometry \citep{Jaskot2019}, which we associate with $a_{\text{asym}}$. In good agreement with this scenario, the \ion{Si}{ii}$^{*}$ EW in the low-$a_{\text{asym}}$ stack is consistent with zero, indicating that a relatively low number of dense molecular clouds imply $f_{\text{cov}} < 1$.

    This unified picture of the ISM in the context of LyC escape is also supported by high-resolution spectroscopic observations of known LyC leakers at $z \sim 2-4$. Most of these galaxies including the Sunburst Arc \citep[$z=2.4$,][]{Rivera-Thorsen2017}, Ion2 \citep[$z=3.2$,][]{Vanzella2015} and Ion3 \citep[$z\approx4.0$,][]{Vanzella2018} show complex \lya{} profiles, often with a central peak at systemic redshift in addition to conventional red and blue peaks \citep[see also][]{Naidu2022}. The simultaneous observation of \lya{} photons that travel through the ISM with and without resonant scattering is naturally explained by the model where the ``picket-fence'' and the density bound geometries are mixed together as described above.
    
    \subsection[C IV: the ISM opacity]{\ion{C}{iv}: the ISM opacity}
    \label{sec:ism_opacity}
    
    In addition to being a tracer of hard radiation fields similarly to the \ion{He}{ii} line, \ion{C}{iv} undergoes resonant scattering in the ISM, probing the density of high-ionisation gas. This makes nebular \ion{C}{iv} emission a proxy for both the production \textit{and} escape of LyC \citep[e.g.][]{Berg2019}. Thanks to these unique properties, the \ion{C}{iv} line has the potential to become one of the most reliable tools for identification of LyC leakers. We note that the same information about the LyC production and escape is also imprinted in \lya{} emission, with the only difference being that the \lya{} profile is sensitive to the column density of neutral gas, whereas the \ion{C}{iv} profile is sensitive to the column density of high-ionisation gas. However, the \lya{} transmission declines rapidly at $z > 6$ due to the increasing opacity of the neutral IGM \citep[e.g.][]{Gronke2021}, which emphasises the importance of \ion{C}{iv} as a standalone probe for $f_{\text{esc}}$ in the EoR.
    
    The LyC diagnostics based on \ion{C}{iv} usually involves comparing it with a reference nebular emission line unaffected by resonant scattering. In particular, recent studies have explored the relationship between LyC leakage and the $\ion{C}{iv}/\ion{C}{iii]}$ ratio. Based on the spectroscopic observations of eight local ($z \sim 0.3-0.4$) SFGs with LyC measurements, \citet{Schaerer2022} proposed the following criterion for classifying a galaxy as a strong LyC leaker ($f_{\text{esc}} > 0.1$): $\ion{C}{IV}/\ion{C}{III]}>0.75$. At higher redshifts ($z = 3.1 - 4.6$), \citet{Saxena2022} found that three out of five \ion{C}{iv} emitters with $f_{\text{esc}} \approx 0.05 - 0.3$ inferred from the LIS absorption lines have similar values of $\ion{C}{IV}/\ion{C}{III]}$ ($\gtrsim 0.75$).
    
    \autoref{fig:civ_ciii} shows $\ion{C}{iv}/\ion{C}{iii]}$ measured from our stacked spectra of LAEs. We find that $\ion{C}{IV}/\ion{C}{III]} \gtrsim 0.75$ in \textit{all} the stacks of LyC-leaker candidates (low-$\upsilon_{\mathrm{sep}}$, FWHM$_{\text{Ly}\alpha}$ Q1 and Q2, and low-$a_{\text{asym}}$), in good agreement with the results from \citet{Schaerer2022}. Conversely, \ion{C}{IV}/\ion{C}{III]} is lower than $\sim$~0.4 in the stacks of potential non-leakers (high-$\upsilon_{\mathrm{sep}}$, FWHM$_{\text{Ly}\alpha}$ Q3 and Q4, and high-$a_{\text{asym}}$). The most striking difference can be seen between the FWHM$_{\text{Ly}\alpha}$ quartiles, where the $\ion{C}{iv}$ to $\ion{C}{iii]}$ ratio drops down from $\ion{C}{iv}/\ion{C}{iii]} = 1.0^{+0.3}_{-0.2}$ (FWHM$_{\text{Ly}\alpha}$, Q1) to zero (FWHM$_{\text{Ly}\alpha}$, Q4).
    
    We argue that the observed \ion{C}{IV}/\ion{C}{III]} ratios in the stacks of potential leakers result not only from more intense ionising radiation fields, but also from the increased transparency of the ISM. First, we compare the \ion{C}{IV} and \ion{He}{ii} emission in the high-$\upsilon_{\mathrm{sep}}$ and -FWHM$_{\text{Ly}\alpha}$ stacks (\autoref{fig:line_ews}, panels f1-f2, g1-g2). We find that the \ion{C}{IV} EW is consistent with zero but the \ion{He}{ii} EW is not, even though the ionisation energy of \ion{C}{IV} (47.9~eV) is lower than that of \ion{He}{ii} (54.4~eV). The absence of \ion{C}{IV} emission from potential non-leakers thus indicates that \ion{C}{IV} experiences a large optical depth from high-ionisation gas. Second, we visually inspect the \ion{C}{IV} velocity offsets. We find that \ion{C}{IV} in the stacks with potential leakers is at the systemic velocity, while in the case of non-leakers, the \ion{C}{IV} line is generally redshifted (\autoref{fig:peaksep_stacks}, \autoref{fig:fwhm_stacks} and \autoref{fig:asym_stacks}). The non-zero velocity offsets are a clear signature of resonant scattering of \ion{C}{IV} by high-ionisation gas, which ultimately leads to a decrease in the observed \ion{C}{IV}/\ion{C}{III]} ratio. Therefore, we conclude that the ISM opacity effects are likely to play an important role in modulating the observed \ion{C}{IV} emission.

\section{Summary}
\label{sec:summary}

    In this work, we select LyC-leaker candidates from a sample of 1422 MUSE LAEs using theoretically- and empirically-motivated criteria based on the \lya{} peak separation ($\upsilon_{\text{sep}}$), full width at half maximum (FWHM$_{\text{Ly}\alpha}$) and asymmetry parameter ($a_{\text{asym}}$). We perform spectral stacking and obtain high-S/N detections of rest-frame UV emission and absorption lines containing valuable information on the ionising properties and the physical conditions in the ISM of our galaxies. By comparing the stacked spectra of potential LyC leakers and non-leakers, we find the following:
    
    \begin{itemize}
        \item The stacks of LyC-leaker candidates generally show strong nebular emission, revealing an extreme ionisation state of the ISM (\autoref{sec:ionising_budget}). This highlights the importance of young, metal-poor stars in creating low-column-density channels in the ISM through which LyC photons escape.
        \item Strong \ion{He}{ii} emission ($\text{EW} \sim 1-3$~\r{A}) is typical among LyC-leaker candidates, implying a significant production rate of ionising photons with energies $>54.4$~eV (\autoref{sec:spectral_hardness}). While elevated $\xi_{\text{ion}}$ is likely to play an important role in boosting the \ion{He}{ii} EW, lower $\ion{C}{iii]}/\ion{He}{ii}$ and $\ion{O}{iii]}/\ion{He}{ii}$ in the stacks of potential LyC leakers suggest that stronger \ion{He}{ii} might also be a consequence of a harder ionising spectrum possibly associated with higher $f_{\text{esc}}$.
        \item The LIS absorption lines are generally weaker for the low-$a_{\text{asym}}$ stacks, suggesting that the asymmetry of the \lya{} line profile depends on the distribution of high-column-density gas and hence traces $f_{\text{cov}}$ in the framework of the ``picket-fence'' model of the ISM. Conversely, the $\upsilon_{\text{sep}}$ and FWHM$_{\text{Ly}\alpha}$ stacks demonstrate a scatter of the LIS absorption line depths, possibly indicating that these \lya{} properties are more sensitive to the density-bounded geometry. Despite having a different impact on the \lya{} line profile, together, the ``picket-fence'' and the density-bounded ISM models provide a coherent physical picture of LyC escape (\autoref{sec:ism_geometry}).
        \item \ion{Si}{ii}$^*$ emission is detected in \textit{all} the $\upsilon_{\text{sep}}$ and FWHM$_{\text{Ly}\alpha}$ stacks, implying the presence of neutral gas off the LOS even in potential LyC leakers (\autoref{sec:ism_geometry}). This indicates that LyC photons escape from a highly anisotropic ISM through ionised channels surrounded by a high-column-density medium.
        \item High $\ion{C}{iv}/\ion{C}{iii]}$ ratios ($>0.75$) are common among the LAEs with potentially high $f_{\text{esc}}$ ($\gtrsim 0.1$; \autoref{sec:ism_opacity}). The $\ion{C}{iv}$ line profile and the comparison between the EWs of $\ion{C}{iv}$ and $\ion{He}{ii}$ indicate that elevated $\ion{C}{iv}/\ion{C}{iii]}$ partly arises from the ISM opacity effects related to LyC escape.
    \end{itemize}

    Our work suggests that the synergy between the extreme ionising fields produced by young stellar populations and the patchy ISM riddled with low-column-density channels creates an ideal physical environment for efficient escape of LyC photons. The $\ion{C}{iv}/\ion{C}{iii]}$ ratio provides the best illustration to our conclusions, being sensitive to both the ionising photon production efficiency and the opacity of the ISM (\autoref{fig:civ_ciii}). Importantly, unlike the \lya{} line, $\ion{C}{iv}$ is not affected by changes in the \ion{H}{i} content of the IGM which makes it a very promising tool for identification and analysis of LyC-leaking galaxies even in the EoR. Calibrating the relationship between $\ion{C}{iv}/\ion{C}{iii]}$ and $f_{\text{esc}}$ at high redshift ($z \gtrsim 3$) will likely become possible in the future thanks to blue-optimised integral-field spectrographs such as BlueMUSE \citep{Richard2019}, which will probe LyC emission from sources at $z \sim 3-4$. Subsequently, $\ion{C}{iv}/\ion{C}{iii]}$ and other spectral properties of SFGs indicative of high $f_{\text{esc}}$ could be measured at $z \gtrsim 6$ with \textit{JWST} and the new generation spectrographs such as MOONS/VLT, PFS/Subaru or HARMONI/ELT, opening new paths for indirect identification of ionising sources in the EoR.

\section*{Acknowledgements}

We thank the anonymous referee for the constructive feedback that helped to improve the manuscript. We thank Michael Maseda, Daniel Schaerer, Charlotte Simmonds and Rashmi Gottumukkala for useful comments and productive discussions. We also thank the organisers and participants of the 24$^{\text{th}}$ MUSE Science Busy Week in Leiden. IGK acknowledges an Excellence Master Fellowship granted by the Faculty of Science of the University of Geneva. This work has received funding from the Swiss State Secretariat for Education, Research and Innovation (SERI) under contract number MB22.00072, as well as from the Swiss National Science Foundation (SNSF) through project grant 200020\_207349 and SNSF Professorship grant 190079. The Cosmic Dawn Center (DAWN) is funded by the Danish National Research Foundation under grant No.\ 140. This paper is based on observations collected at the European Organisation for Astronomical Research in the Southern Hemisphere under ESO programmes 094.A-0289(B), 095.A-0010(A), 096.A-0045(A), 096.A-0045(B), 094.A-0205, 095.A-0240, 096.A-0090, 097.A-0160, and 098.A-0017. We made extensive use of several open-source software packages and we are thankful to the respective authors for sharing their work: \textsc{numpy} \citep{Numpy2020}, \textsc{astropy} \citep{Astropy2022}, \textsc{matplotlib} \citep{Matplotlib2007}, \textsc{ipython} \citep{IPython2007} and \textsc{topcat} \citep{Topcat2005}.

\section*{Data Availability}

Raw MUSE data cubes are available from the ESO Science Archive Facility\footnote{\url{http://archive.eso.org}}. The MUSE-Wide data from the first data release, which covers 44 fields in GOODS-S/CDFS, are available on the MUSE-Wide project page\footnote{\url{https://musewide.aip.de}}. The MUSE HUDF data from the second data release are available via the Advanced MUSE Data products (AMUSED) web interface\footnote{\url{https://amused.univ-lyon1.fr}}. The stacked spectra in a machine-readable format and the spectral line measurements will be shared upon request to the corresponding author.



\bibliographystyle{mnras}
\bibliography{main}



\appendix

\onecolumn

\makeatletter
\renewcommand{\fps@figure}{htbp}
\renewcommand{\fps@table}{htbp}
\makeatother

\begin{landscape}
    \section{Equivalent widths of rest-frame UV lines detected in the stacked spectra of MUSE LAEs}

    \begin{table}
        \caption{EWs of rest-frame UV lines detected in the stacked spectra of MUSE LAEs.}
        \label{tab:line_ews}
        \renewcommand{\arraystretch}{1.3}
        \begin{tabular}{llrrrrrrrrrrr}
            Sample & $\Sigma$ & Lyman~$\alpha$ & \ion{O}{I}+\ion{Si}{II}$\lambda\lambda$1302,4 & \ion{C}{II}$\lambda$1335 & \ion{Si}{II}$\lambda$1527 & \ion{Si}{II}$^{*}\lambda$1533 & \ion{C}{IV}$\lambda\lambda$1548,51 & \ion{He}{II}$\lambda1640$ & \ion{O}{III]}$\lambda$1666 & \ion{[Si}{III]}$\lambda$1883 & \ion{C}{III]}$\lambda\lambda$1907,9 \\
            (1) & (2) & (3) & (4) & (5) & (6) & (7) & (8) & (9) & (10) & (11) & (12)\\
            \hline
            \hline
            All & 1422 & $83.6_{-1.5}^{+1.3}$ & $-0.4_{-0.1}^{+0.1}$ & $-0.8_{-0.2}^{+0.2}$ & $-0.6_{-0.2}^{+0.2}$ & $1.2_{-0.2}^{+0.2}$ & $2.2_{-0.5}^{+0.5}$ & $1.1_{-0.3}^{+0.3}$ & $1.9_{-0.3}^{+0.3}$ & $1.0_{-0.5}^{+0.5}$ & $5.7_{-0.7}^{+0.8}$ \\
            \hline
            \hline
            $\upsilon_{\mathrm{sep}} < 463$~km~s$^{-1}$ & 165 & $100.2_{-4.7}^{+5.3}$ & $-0.7_{-0.8}^{+0.7}$ & $-0.7_{-0.8}^{+0.8}$ & $-1.9_{-0.6}^{+0.6}$ & $1.0_{-0.5}^{+0.5}$ & $4.0_{-1.1}^{+1.2}$ & $2.9_{-1.1}^{+1.0}$ & $2.3_{-0.8}^{+0.9}$ & $0.9_{-0.4}^{+0.8}$ & $8.1_{-2.2}^{+2.3}$ \\
            $\upsilon_{\mathrm{sep}} > 463$~km~s$^{-1}$ & 164 & $61.9_{-2.4}^{+2.4}$ & $-0.8_{-0.3}^{+0.3}$ & $-1.2_{-0.2}^{+0.3}$ & $0.7_{-0.5}^{+0.5}$ & $1.4_{-0.3}^{+0.3}$ & $0.6_{-0.7}^{+0.8}$ & $1.0_{-0.3}^{+0.3}$ & $1.3_{-0.4}^{+0.4}$ & $1.6_{-0.6}^{+0.7}$ & $5.1_{-1.0}^{+0.9}$ \\
            \hline
            FWHM$_{\mathrm{Ly}\alpha}$ (Q1) & 241 & $162.5_{-10.1}^{+9.3}$ & $0.5_{-1.4}^{+1.3}$ & $-1.8_{-0.9}^{+0.7}$ & $-1.6_{-0.7}^{+0.5}$ & $5.5_{-1.8}^{+1.9}$ & $14.1_{-2.3}^{+2.5}$ & $3.4_{-1.2}^{+1.1}$ & $3.8_{-1.0}^{+1.8}$ & $3.1_{-1.1}^{+1.3}$ & $6.9_{-1.4}^{+1.8}$ \\
            FWHM$_{\mathrm{Ly}\alpha}$ (Q2) & 241 & $144.6_{-5.5}^{+5.5}$ & $0.7_{-0.9}^{+1.0}$ & $0.3_{-1.0}^{+1.0}$ & $-1.1_{-0.6}^{+0.4}$ & $1.9_{-0.6}^{+0.7}$ & $6.4_{-1.4}^{+1.4}$ & $1.3_{-0.5}^{+0.6}$ & $5.1_{-1.0}^{+0.8}$ & $1.1_{-0.5}^{+0.6}$ & $7.1_{-2.4}^{+1.5}$ \\
            FWHM$_{\mathrm{Ly}\alpha}$ (Q3) & 241 & $105.4_{-5.6}^{+5.2}$ & $-0.6_{-0.2}^{+0.3}$ & $-1.2_{-0.4}^{+0.5}$ & $-1.8_{-0.8}^{+0.8}$ & $1.3_{-0.5}^{+0.5}$ & $1.6_{-0.6}^{+0.6}$ & $1.6_{-0.5}^{+0.6}$ & $2.1_{-0.6}^{+0.7}$ & $2.4_{-0.9}^{+1.0}$ & $7.8_{-1.4}^{+1.3}$ \\
            FWHM$_{\mathrm{Ly}\alpha}$ (Q4) & 241 & $51.2_{-1.9}^{+1.8}$ & $-0.8_{-0.2}^{+0.2}$ & $-0.9_{-0.2}^{+0.2}$ & $-1.0_{-0.3}^{+0.3}$ & $0.6_{-0.4}^{+0.4}$ & $-0.3_{-0.5}^{+0.5}$ & $0.6_{-0.3}^{+0.3}$ & $0.7_{-0.2}^{+0.3}$ & $1.2_{-0.4}^{+0.5}$ & $2.8_{-0.9}^{+0.7}$ \\
            \hline
            a$_{\mathrm{asym}} < 0.19$ & 277 & $87.3_{-2.8}^{+2.9}$ & $-0.2_{-0.3}^{+0.4}$ & $-0.9_{-0.3}^{+0.2}$ & $-0.7_{-0.4}^{+0.4}$ & $0.3_{-0.4}^{+0.4}$ & $2.6_{-0.6}^{+0.6}$ & $1.4_{-0.4}^{+0.3}$ & $2.4_{-0.5}^{+0.5}$ & $1.5_{-0.5}^{+0.4}$ & $5.4_{-1.0}^{+0.7}$ \\
            a$_{\mathrm{asym}} > 0.19$ & 277 & $68.8_{-2.2}^{+2.1}$ & $-0.6_{-0.2}^{+0.2}$ & $-0.9_{-0.3}^{+0.3}$ & $-1.1_{-0.3}^{+0.3}$ & $1.2_{-0.4}^{+0.5}$ & $1.2_{-0.6}^{+0.5}$ & $0.8_{-0.3}^{+0.3}$ & $2.2_{-0.8}^{+0.6}$ & $-0.8_{-0.9}^{+1.0}$ & $5.7_{-1.2}^{+1.5}$ \\
            \hline
            \hline
            $\mathrm{EW}_{\mathrm{Ly}\alpha}$ (Q1) & 262 & $34.3_{-1.5}^{+1.5}$ & $-0.3_{-0.4}^{+0.4}$ & $-1.0_{-0.2}^{+0.2}$ & $-0.7_{-0.3}^{+0.3}$ & $1.6_{-0.5}^{+0.5}$ & $0.7_{-0.5}^{+0.4}$ & $0.4_{-0.2}^{+0.2}$ & $0.6_{-0.3}^{+0.3}$ & $0.8_{-0.4}^{+0.6}$ & $1.2_{-0.8}^{+1.9}$ \\
            $\mathrm{EW}_{\mathrm{Ly}\alpha}$ (Q2) & 262 & $70.2_{-2.8}^{+2.5}$ & $-0.7_{-0.3}^{+0.3}$ & $-0.7_{-0.4}^{+0.3}$ & $-1.2_{-0.4}^{+0.4}$ & $0.7_{-0.3}^{+0.4}$ & $1.9_{-0.5}^{+0.4}$ & $0.2_{-0.7}^{+0.7}$ & $2.3_{-0.6}^{+0.7}$ & $1.3_{-0.4}^{+0.5}$ & $3.4_{-0.7}^{+0.7}$ \\
            $\mathrm{EW}_{\mathrm{Ly}\alpha}$ (Q3) & 262 & $101.3_{-3.7}^{+3.8}$ & $-0.9_{-0.4}^{+0.3}$ & $-1.3_{-0.5}^{+0.5}$ & $-1.3_{-0.5}^{+0.6}$ & $2.0_{-0.7}^{+0.6}$ & $2.2_{-0.7}^{+1.2}$ & $0.5_{-0.3}^{+0.4}$ & $2.5_{-0.6}^{+0.6}$ & $0.9_{-0.4}^{+0.5}$ & $9.9_{-2.2}^{+2.0}$ \\
            $\mathrm{EW}_{\mathrm{Ly}\alpha}$ (Q4) & 262 & $176.6_{-4.7}^{+4.1}$ & $1.5_{-1.0}^{+1.0}$ & $-1.5_{-0.6}^{+0.5}$ & $-0.9_{-1.2}^{+1.1}$ & $0.1_{-1.0}^{+1.1}$ & $6.4_{-1.8}^{+1.8}$ & $1.5_{-0.6}^{+0.7}$ & $4.1_{-1.2}^{+1.4}$ & $3.9_{-1.6}^{+3.4}$ & $25.0_{-5.9}^{+4.9}$ \\
            \hline
            $L_{\mathrm{Ly}\alpha}$ (Q1) & 356 & $91.1_{-1.9}^{+2.2}$ & $-1.7_{-0.9}^{+0.8}$ & $1.3_{-0.9}^{+0.9}$ & $0.9_{-1.0}^{+0.9}$ & $2.1_{-0.5}^{+0.6}$ & $3.5_{-1.1}^{+1.0}$ & $1.0_{-0.7}^{+0.7}$ & $2.4_{-0.6}^{+0.6}$ & $1.7_{-0.9}^{+1.2}$ & $4.9_{-1.3}^{+1.3}$ \\
            $L_{\mathrm{Ly}\alpha}$ (Q2) & 355 & $86.5_{-2.4}^{+2.4}$ & $-0.8_{-0.2}^{+0.2}$ & $-1.0_{-0.5}^{+0.3}$ & $-1.1_{-0.7}^{+0.7}$ & $0.7_{-0.4}^{+0.4}$ & $1.6_{-0.6}^{+0.6}$ & $0.6_{-0.2}^{+0.5}$ & $2.6_{-0.6}^{+0.6}$ & $0.8_{-0.4}^{+0.4}$ & $5.8_{-1.2}^{+1.7}$ \\
            $L_{\mathrm{Ly}\alpha}$ (Q3) & 355 & $86.2_{-2.0}^{+2.3}$ & $-0.9_{-0.3}^{+0.4}$ & $-1.0_{-0.3}^{+0.3}$ & $-0.7_{-0.4}^{+0.2}$ & $1.5_{-0.5}^{+0.5}$ & $3.2_{-0.8}^{+1.1}$ & $0.8_{-0.2}^{+0.2}$ & $1.1_{-0.5}^{+0.6}$ & $2.5_{-0.9}^{+1.3}$ & $7.3_{-1.1}^{+0.8}$ \\
            $L_{\mathrm{Ly}\alpha}$ (Q4) & 355 & $91.6_{-2.1}^{+2.1}$ & $0.2_{-0.5}^{+0.4}$ & $-0.8_{-0.5}^{+0.3}$ & $-1.2_{-0.4}^{+0.3}$ & $1.0_{-0.4}^{+0.4}$ & $2.3_{-0.5}^{+0.6}$ & $1.2_{-0.5}^{+0.5}$ & $2.4_{-0.4}^{+0.4}$ & $1.1_{-0.5}^{+0.6}$ & $4.4_{-1.4}^{+1.2}$ \\
            \hline
            \hline
        \end{tabular}
        
        \textbf{Notes.} Columns are as follows: (1) sample description, (2) sample size, (3) \lya{} EW, (4-6) EWs of ISM absorption lines (\ion{O}{I}$\lambda$1302$+$\ion{Si}{II}$\lambda$1304, \ion{C}{II}$\lambda$1335 and \ion{Si}{II}$\lambda$1527), (7) EW of the \ion{Si}{II}$^{*}\lambda$1533 fine-structure emission line, (8-12) EWs of high-ionisation nebular emission lines (\ion{C}{IV}$\lambda\lambda$1548,1551, \ion{He}{II}$\lambda1640$, \ion{O}{III]}$\lambda$1666, \ion{[Si}{III]}$\lambda$1883 and \ion{[C}{III]}$\lambda$1907$+$\ion{C}{III]}$\lambda$1909). All EWs are given for the rest-frame spectra and expressed in \r{A}.
    \end{table}

\end{landscape}

\section{Stacked spectra of potential LyC leakers and non-leakers}
\label{appendix:stacks}

    \subsection[Subsamples of different Lyman-alpha FWHMs]{Subsamples of different \lya{}~FWHMs}
    \begin{figure*}
        \includegraphics[width=\linewidth]{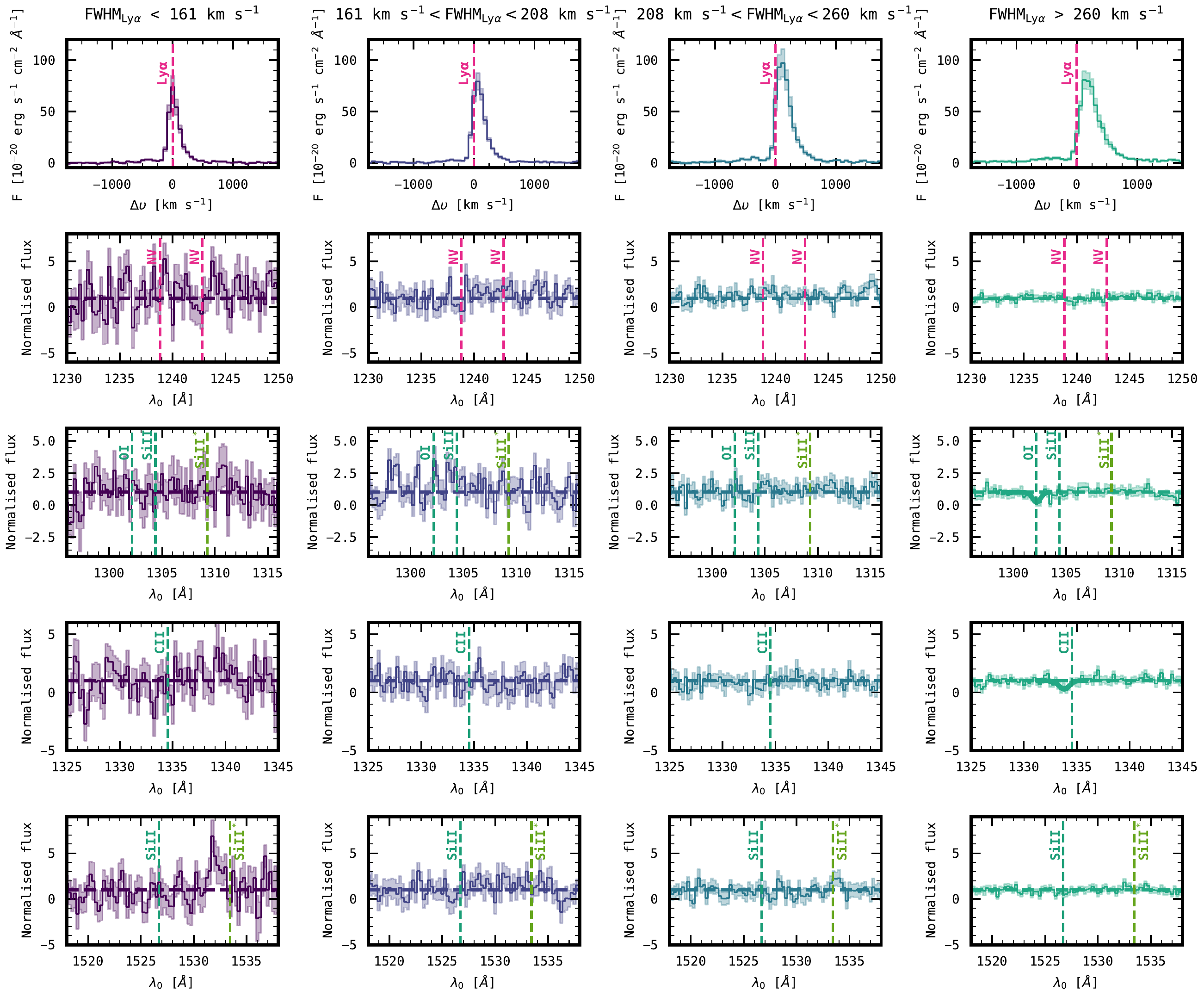}
        \caption{Similar to \autoref{fig:peaksep_stacks}, but for the subsamples of LAEs with different \lya{}~FWHMs.}
        \label{fig:fwhm_stacks}
    \end{figure*}
    
    \begin{figure*}
        \includegraphics[width=\linewidth]{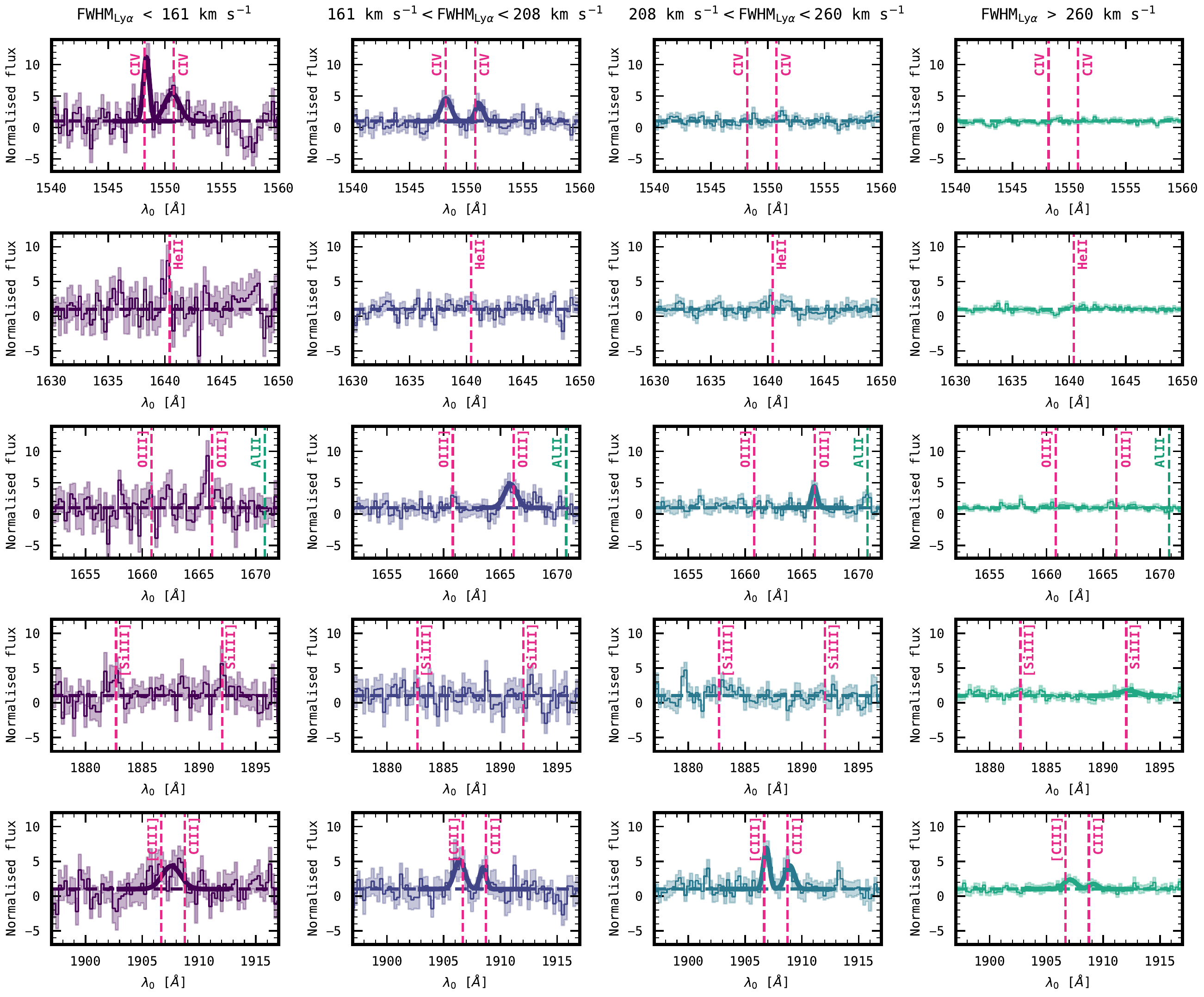}
        \contcaption{}
    \end{figure*}
    
    \clearpage
    \subsection[Subsamples of different Lyman-alpha asymmetry parameters]{Subsamples of different \lya{}~asymmetry parameters}
    \begin{figure*}
        \includegraphics[width=\linewidth]{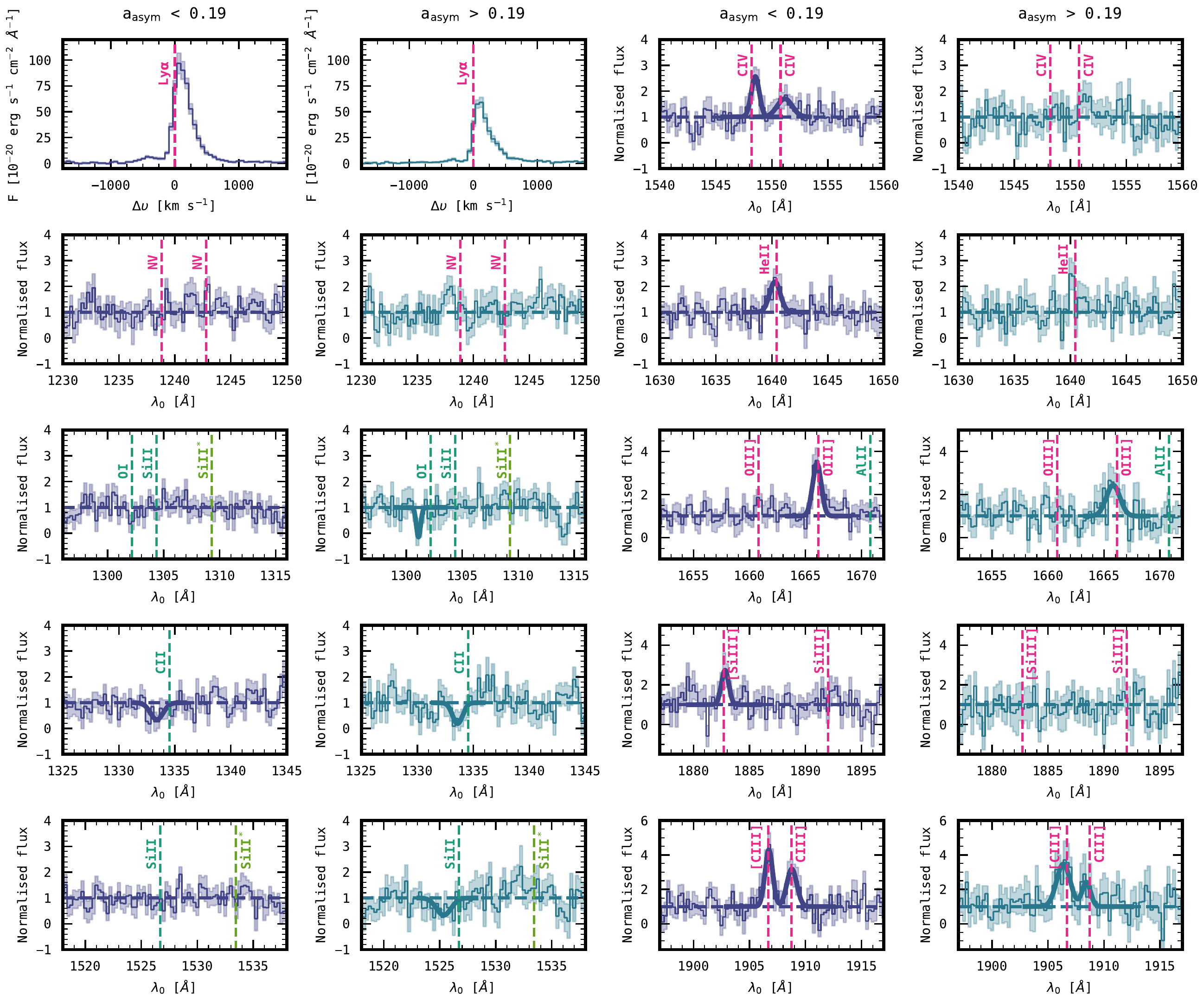}
        \caption{Similar to \autoref{fig:peaksep_stacks}, but for the subsamples of LAEs with low ($< 0.19$; blue) and high ($> 0.19$; green) \lya{}~asymmetry parameters.}
        \label{fig:asym_stacks}
    \end{figure*}

\clearpage
\section[Equivalent widths of rest-frame UV lines for different Lyman-alpha EWs and luminosities]{Equivalent widths of rest-frame UV lines for different \lya{}~EWs and luminosities}
\label{appendix:ews}
    
    \begin{figure*}
        \includegraphics[width=0.67\linewidth]{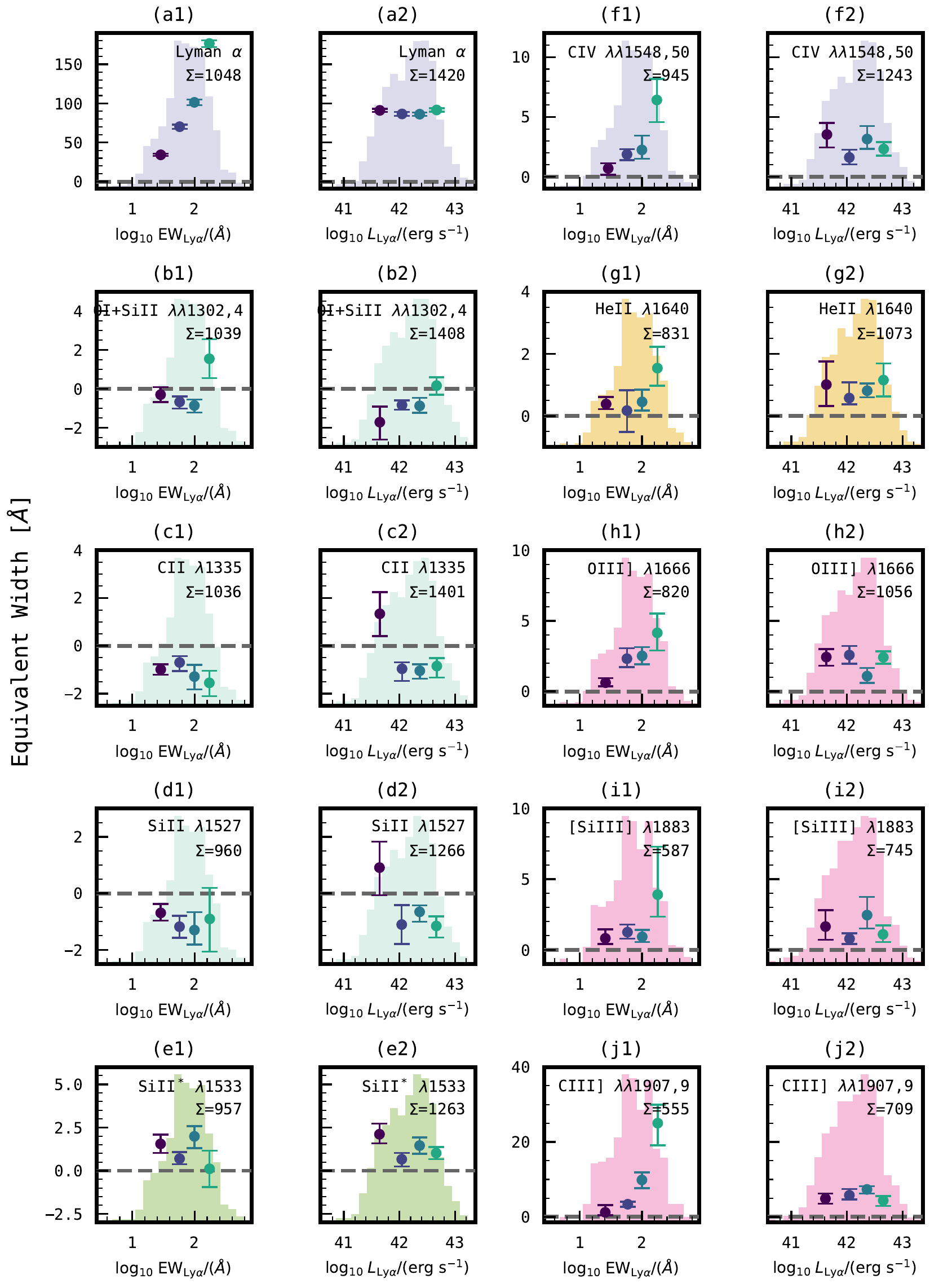}
        \caption{Similar to \autoref{fig:line_ews}, but for the subsamples of LAEs with different \lya{}~EWs and luminosities.}
        \label{fig:line_ews_app}
    \end{figure*}


\bsp	
\label{lastpage}
\end{document}